\documentclass[aps,pra,twocolumn,superscriptaddress]{revtex4-2}
\usepackage{amsmath}    
\usepackage{graphicx}
\usepackage{booktabs}
\usepackage{xcolor,soul}
\usepackage{amssymb}
\usepackage{mathdots}
\definecolor{darkblue}{HTML}{3771C8}
\usepackage[%
    colorlinks=true,
    pdfborder={0 0 0},
    linkcolor=darkblue,
    citecolor = darkblue, 
    urlcolor = darkblue
]{hyperref}
\usepackage[capitalize,nameinlink]{cleveref}
\usepackage[normalem]{ulem}

\newcommand{\ef}{|e\rangle\xrightarrow[]{}|f\rangle}
\newcommand{\gee}{|g\rangle\xrightarrow[]{}|e\rangle}
\newcommand{\ke}{\sqrt{\dfrac{\kappa_\text{em}}{2}}}
\newcommand{\nth}{\Bar{n}_\text{th}}
\newcommand{\exc}[1]{\langle #1 \rangle}
\newcommand{\tp}{\tau_\text{P}}

\newcommand{\Gf}{\Gamma_\text{1D}^\text{f}}
\newcommand{\Gb}{\Gamma_\text{1D}^\text{b}}

\begin{document}
\title{Resonance fluorescence of a chiral artificial atom}

\author{Chaitali Joshi}
    \email[These authors contributed equally.]{}
    \affiliation{Moore Laboratory of Engineering, California Institute of Technology, Pasadena, California 91125}
    \affiliation{Institute for Quantum Information and Matter, California Institute of Technology, Pasadena, California 91125}

\author{Frank Yang}
    \email[These authors contributed equally.]{}
    \affiliation{Moore Laboratory of Engineering, California Institute of Technology, Pasadena, California 91125}
    \affiliation{Institute for Quantum Information and Matter, California Institute of Technology, Pasadena, California 91125}

\author{Mohammad Mirhosseini}
    \email[mohmir@caltech.com; http://qubit.caltech.edu]{}
    \affiliation{Moore Laboratory of Engineering, California Institute of Technology, Pasadena, California 91125}
    \affiliation{Institute for Quantum Information and Matter, California Institute of Technology, Pasadena, California 91125}

\date{\today} 

\begin{abstract}We demonstrate a superconducting artificial atom with strong unidirectional coupling to a microwave photonic waveguide. Our artificial atom is realized by coupling a transmon qubit to the waveguide at two spatially separated points with time-modulated interactions. Direction-sensitive interference arising from the parametric couplings in our scheme results in a non-reciprocal response, where we measure a forward/backward ratio of spontaneous emission exceeding 100. We verify the quantum nonlinear behavior of this artificial chiral atom by measuring the resonance fluorescence spectrum under a strong resonant drive and observing well-resolved Mollow triplets. Further, we demonstrate chirality for the second transition energy of the artificial atom and control it with a pulse sequence to realize a qubit-state-dependent non-reciprocal phase on itinerant photons. Our demonstration puts forth a superconducting hardware platform for the scalable realization of several key functionalities pursued within the paradigm of chiral quantum optics, including quantum networks with all-to-all connectivity, driven-dissipative stabilization of many-body entanglement, and the generation of complex non-classical states of light.

\end{abstract}


\maketitle

\subsection*{Introduction} Chiral light-matter interfaces have been long studied in quantum optics \cite{10.1103/physrevlett.70.2269,10.1103/physrevlett.70.2273} and promise a myriad of potential opportunities for developing quantum networks with long-range connectivity \cite{ Cirac:1997is,10.1103/physrevlett.118.133601,Xiang:2017jd}, and generating novel many-body entangled states of light \cite{wan2021, pichler2017} and matter \cite{stannigel2012, pichler2015}. Light-matter interaction is said to be \emph{chiral} when the scattering of a photon from an atom depends strongly on the photon's propagation direction in a one-dimensional (1D) waveguide \cite{lodahl2017}. This breaking of the symmetry of atom-waveguide coupling to the right and left propagating modes gives rise to a range of unique phenomena. For example, a resonant photon impinging on a chiral atom strongly coupled to a 1D waveguide acquires a non-reciprocal $\pi$ phase shift conditioned on the state of the atom. This remarkable effect can be exploited to realize entangling gates between distant stationary qubits mediated by itinerant photons \cite{guimond2020, mahmoodian2016a}. In the paradigm of waveguide quantum electrodynamics (QED), coupling several chiral two-level systems to a common waveguide results in novel collective spin dynamics and the formation of exotic non-equilibrium phases of entangled spin clusters \cite{pichler2015}. Conversely, complex non-classical states of light such as multidimensional cluster states and Fock states can be generated efficiently using protocols that rely on deterministic chiral atom-photon interactions  \cite{pichler2017, yang2022}.  

Chiral atom-photon interfaces have been realized in the optical domain by coupling atoms and solid-state quantum emitters to nanophotonic structures, where the strong confinement of light results in the locking of the local polarization of a photon to its direction of propagation \cite{sollner2015, lodahl2017, lefeber2015, petersen2014, sayrin2015, scheucher2016}. Despite remarkable progress in these systems, achieving strong unidirectional coupling with a chain of emitters remains challenging due to the relative weakness of interactions in the case of single atoms and the environment-induced frequency disorder of solid-state emitters. More recently, artificial atoms based on superconducting qubits have emerged as a powerful platform for waveguide QED in the microwave domain. These systems offer control over individual emitters and their coupling to the environment, as well as the ability to tailor the dispersion of the electromagnetic modes in waveguides. Additionally, the relatively large wavelength of radiation at the GHz band allows for the precise placement of atoms along a waveguide to control photon-mediated interactions and collective dissipation \cite{vanloo2013, Mirhosseini2019May}. These advantages have enabled several demonstrations of waveguide QED phenomena with superconducting artificial atoms, including resonance fluorescence \cite{Astafiev2010Feb, Hoi2013Feb, Hoi2015Dec}, Dicke super- and sub-radiance\cite{vanloo2013,Mirhosseini2019May, zanner2022}, formation of qubit-photon bound states \cite{liu2017Jan, sundaresan2019} and the realization of long-range waveguide-mediated coupling for many-body quantum simulations \cite{zhang2022}. Despite this rapid progress, studying chiral quantum optics with superconducting qubits remains challenging due to the lack of an efficient unidirectional interface for microwave photons.

Nonreciprocal transport of microwave photons is possible using devices based on ferro/ferrimagnetic materials, which break Lorentz reciprocity. Recently, three-dimensional qubit-cavity systems have successfully realized chiral interactions using this approach \cite{owens2022}. However, ferromagnetic devices such as circulators are not suitable for on-chip integration due to their size, large magnetic fields, and typically lossy response. Alternatively, low-loss nonreciprocal components have been realized using synthetic gauge fields \cite{metelmann2015,sliwa2015,chapman2017a, 10.48550/arxiv.2103.07793}. While these experiments demonstrate the non-reciprocal propagation of microwave photons, a simple and scalable approach for realizing on-chip chiral interactions with superconducting qubits remains desirable. More recently, unidirectional emission and absorption of microwave photons have been proposed \cite{10.1103/physrevlett.127.233601,gheeraert2020, guimond2020} and demonstrated \cite{Kannan2022Mar, 10.48550/arxiv.2205.03293,rosario2018} using a pair of entangled qubits. 
However, relying on the interference of the emission from two distinct physical qubits limits the chiral behavior to weak drives where at most a single photon is exchanged with a radiative bath. This diluted quantum nonlinear response forbids a direct realization of strongly driven-dissipative quantum systems. 


%

Here, we experimentally demonstrate a chiral artificial atom consisting of a transmon qubit coupled to a transmission line at two spatially separated points, operating in the so-called \emph{giant-atom} regime \cite{FriskKockum2014Jul,kockum2018a, Vadiraj2021Feb, kannan2020}. The emitted field components from the two coupling points are imparted a relative phase using time-modulated parametric couplings. In this setting, chirality arises from the interference between the two emission pathways resulting from the phase difference from the parametric couplings and the direction-dependent phase delay from propagation in the waveguide. We show highly directional atom-waveguide coupling, with the rate of spontaneous emission to the forward propagating modes exceeding that of backward propagating modes by more than two orders of magnitude. Relying on a single physical qubit as the emission source, our scheme is robust against decoherence and preserves quantum nonlinear response under strong drives. We demonstrate this quantum
nonlinearity using resonance fluorescence measurements
and observe Mollow triplets under a
strong resonant drive. The chiral response is further shown to be continuously tunable and extends to the transmon qubit's second transition ($\ef$). Finally, we use time-dimain control to realize a qubit-state-dependent response to traveling photons in the waveguide. The minimal hardware overhead in our experiment, combined with near-perfect directionality, in situ control of the coupling, and access to higher-order chiral transitions, provide a scalable platform for future studies of driven-dissipative entanglement generation, quantum networks with all-to-all coupling, and cascaded quantum systems with superconducting qubits. 

\subsection*{Design of the artificial atom}

\begin{figure*}[htbp]
\centering
\includegraphics[width=0.95\linewidth]{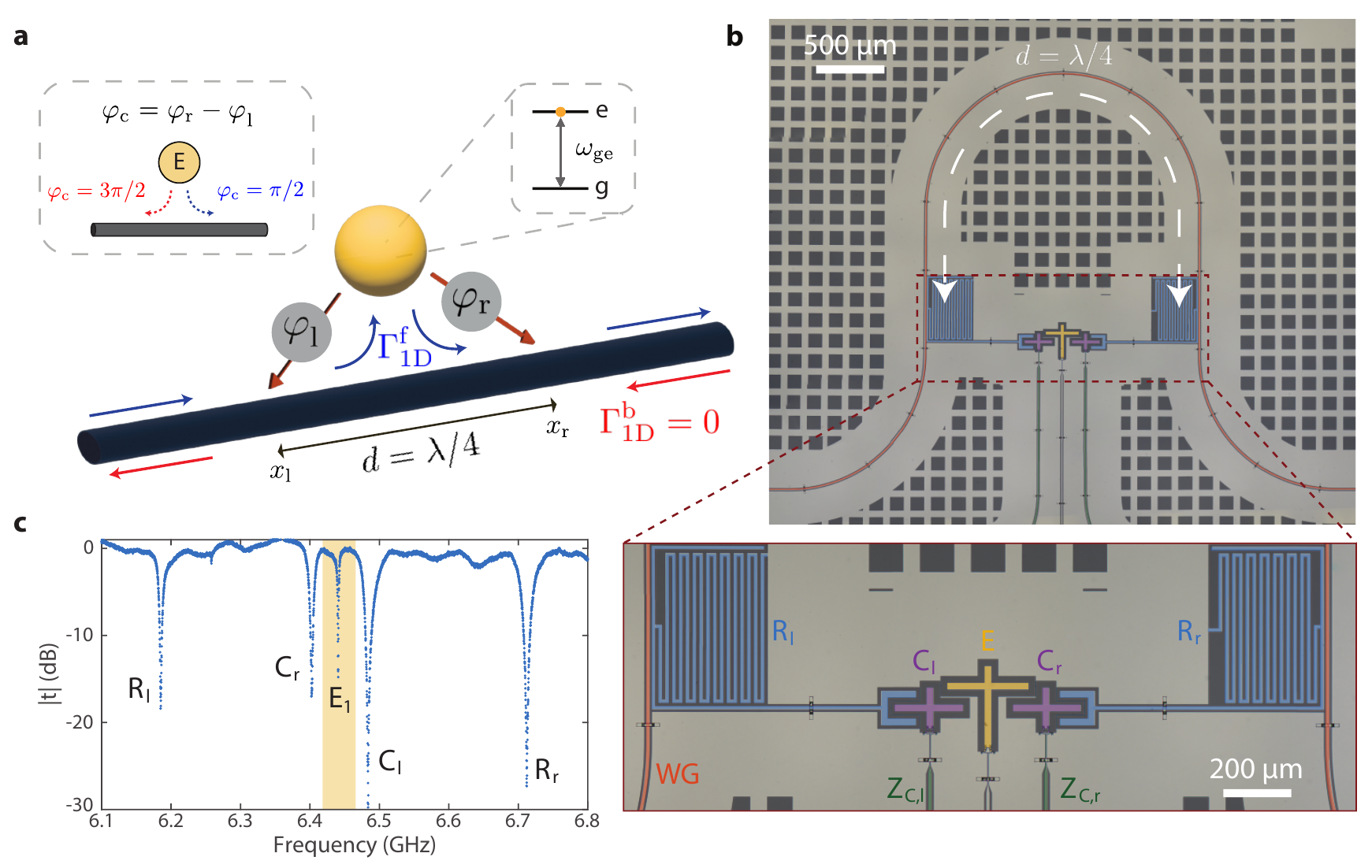}
\caption{\textbf{Chiral atom coupled to a waveguide.} (a) Schematic of chiral atom-waveguide system. The emitter atom couples to the waveguide at two points separated by a distance $d = \lambda/4$. Time-modulated coupling imparts a phase of $\varphi_\mathrm{l (r)}$ at the left (right) point; the relative phase $\varphi_{\mathrm{c}} = \varphi_{\mathrm{r}} - \varphi_{\mathrm{l}}$ tunes interference between the two radiation pathways. $\varphi_{\mathrm{c}} = \pi/2$ results in maximum coupling to forward propagating modes (blue) and no coupling to backward propagating modes (red). Left inset: Varying $\varphi_{\mathrm{c}}$ results in coupling to waveguide modes of opposite directions. (b) Optical image of the fabricated device. The emitter transmon (yellow, E) couples to a microwave co-planar waveguide (orange, WG) at two points separated by $d =$ 4.590 mm. At each point, radiation to the waveguide is mediated by a frequency-selective dissipation port containing a tunable coupler (purple C$_{\mathrm{(l)r}}$) and a filter cavity (blue, R$_{\mathrm{(l)r}}$) directly coupled to the waveguide (orange, WG). A pair of flux bias lines (green, $Z_\mathrm{C,l(r)}$) are used to drive the couplers parametrically. (c) Transmission spectrum $|t| = |\langle\hat{a}^\mathrm{f}_\mathrm{out}\rangle/\langle \hat{a}^\mathrm{f}_\mathrm{in}\rangle|$ of the device under the experimental settings 
used for achieving maximum chirality. Parametric driving of the couplers leads to a visible sideband for the emitter transmon (E$_1$) at the center of the measurement band. C$_{\mathrm{l(r)}}$ (R$_{\mathrm{l(r)}}$) mark the resonances corresponding to the couplers (filter cavities).}
\label{fig:1}
\end{figure*}



As shown in \cref{fig:1}a, our experiment is based on a planar transmon qubit (hereafter called the `emitter') coupled at two locations to an on-chip co-planar waveguide. We use a dissipation port at each coupling point, which is designed to realize a complex coupling rate to the waveguide modes with a well-defined phase ($\varphi_\mathrm{l,r}$ for the left/right ports). The Hamiltonian for this system is given by $\hat{H} = \hat{H}_\mathrm{atom} + \hat{H}_\mathrm{field} + \hat{H}_\mathrm{int}$, where $\hat{H}_\mathrm{atom}/\hbar = \omega_\mathrm{ge}|e\rangle \langle e|$ and $\hat{H}_\mathrm{field}/\hbar = \int dk \omega_k (\hat{a}_{k,\mathrm{f}}^\dagger \hat{a}_{k,{\mathrm{f}}} + \hat{a}^\dagger_{k,\mathrm{b}} \hat{a}_{k,\mathrm{b}})$ is the free Hamiltonian for the atom and the field modes, respectively.  The integral in $\hat{H}_\mathrm{field}$ is performed over positive values of the photonic wavevectors $k$, and the subscripts denote forward ($\mathrm{f}$) and backward ($\mathrm{b}$) propagating modes in the waveguide. The atom-waveguide interaction Hamiltonian can be written as
\begin{align}
    \hat{H}_\mathrm{int}/\hbar =\sum_{i \in \mathrm{l,r}} \int dk [\tilde{g}_{k,i} \hat{\sigma}_{-} (\hat{a}^\dagger_{k,\mathrm{f}} e^{ik x_i} + \hat{a}^\dagger_{k,\mathrm{b}} e^{-ikx_i}) + \mathrm{H.c.}],
    \label{eq:atom-waveguide-hamiltonian}
\end{align}
where $\tilde{g}_{k,\mathrm{l(r)}}$ denotes the complex atom-waveguide coupling strength at the left (right) coupling points, $\hat{\sigma}_{-}$ is the emitter's lowering operator and $x_\mathrm{l(r)}$ are the position of the left (right) coupling point along the waveguide (see \cref{fig:1}a). Assuming a waveguide with linear dispersion, the photonic mode resonant with the $\gee$ transition of the atom (frequency $\omega_\mathrm{ge}$) has a wavevector $k = \omega_\mathrm{ge}/v$, where $v$ is the speed of propagation of the modes in the waveguide. Denoting the atom-waveguide coupling strength for this mode as $\tilde{g}_\mathrm{l(r)}$ at the left (right) coupling points, the decay rate of the atom to the waveguide is given by $\kappa_\mathrm{em, l(r)} = 4 \pi |\tilde{g}_\mathrm{l(r)}|^2 D(\omega_\mathrm{ge})$ where $D(\omega)$ is the density of states in the waveguide \cite{FriskKockum2014Jul}. The emission field of the atom acquires a phase $\varphi_\mathrm{l (r)} = \mathrm{arg} [\tilde{g}_\mathrm{l(r)}]$ at the coupling points. In addition, the distance $d = x_\mathrm{r} - x_\mathrm{l}$ sets the propagation phase $\varphi_\mathrm{WG} = \omega_\mathrm{ge}d/v$ between the two coupling points. When setting $d = \lambda/4$ (where $\lambda = 2\pi v/\omega_\mathrm{ge}$ is the wavelength of the photons), a photon emitted from one coupling point accumulates a $\pi/2$ phase-shift when propagating to the adjacent coupling point. In this situation, setting the relative phase $\varphi_\mathrm{c} = \varphi_\mathrm{r}-\varphi_\mathrm{l}=\pi/2$ results in a chiral response, with the emission from the two ports interfering constructively (destructively) in the forward (backward) direction. Formally, the spatially non-local emitter-waveguide coupling can be described using the SLH formalism (see \cref{appendix:input-output}), leading to a pair of input-output relations for the forward and backward propagating modes.

\begin{align}
\hat{a}^\mathrm{f}_\mathrm{out} & = \hat{a}_\mathrm{in}^\mathrm{f} + (1 + e^{i(\varphi_\mathrm{c}-\varphi_\mathrm{WG})})\sqrt{\frac{\kappa_\mathrm{em}}{2}}\hat{\sigma}_{-}, \\
\hat{a}^\mathrm{b}_\mathrm{out} & =\hat{a}_\mathrm{in}^\mathrm{b} + (1 + e^{i(\varphi_\mathrm{c}+\varphi_\mathrm{WG})})\sqrt{\frac{\kappa_\mathrm{em}}{2}}\hat{\sigma}_{-}.
\label{eq:chirality_conditions} 
\end{align}
Here, $\hat{a}_\mathrm{in}^\mathrm{f(b)}$ is the input field for the forward (backward) propagating mode, $\hat{a}_\mathrm{out}^\mathrm{f(b)}$ is the corresponding output field and we have assumed that the magnitude of the decay rate at the two coupling points is equal and given by $\kappa_\mathrm{em}$. Solving the input-output relations yields the transmission $t = \langle \hat{a}^\mathrm{f}_\mathrm{out} \rangle / \langle \hat{a}^\mathrm{f}_\mathrm{in} \rangle$ and the emitter's rate of spontaneous emission into the forward (backward) direction 
\begin{equation}
\Gamma^\mathrm{f(b)}_\mathrm{1D}/\kappa_\mathrm{em} = 1+ \cos(\varphi_\mathrm{c} \mp \varphi_\mathrm{WG}).
\label{eq:Gamma1D}
\end{equation}
Note that, in principle, any waveguide length permits full suppression of coupling to one waveguide direction (with the exception of $d = n{\lambda}/{2}$, for $n \in \mathbb{Z}$). This maximum chirality condition only coincides with the maximum emitter external decay rate when $d = (2n+1){\lambda}/{4}$, motivating our choice of the waveguide length ($\varphi_\mathrm{WG} = \pi/2$), which results in $\Gamma^\mathrm{b}_\mathrm{1D} = 0$ ($\Gamma^\mathrm{f}_\mathrm{1D} = 2\kappa_\mathrm{em}$) at $\varphi_c = \pi/2$.

To realize complex coupling strengths, we rely on a periodic modulation of the photon hopping rates from the emitter to the waveguide \cite{10.1038/nphys3930}. We achieve this by frequency modulating a coupler device \cite{yan2018a} that is capacitively coupled to the emitter. In this configuration, the coupler is modulated with a sinusoidal flux drive at the frequency $\Delta$, with an amplitude $\epsilon_\mathrm{(l)r}$ and phase $\varphi_\mathrm{l(r)}$, resulting in an effective emitter-waveguide coupling term that picks up the driving phase $\arg[\tilde{g}_\mathrm{l(r)}] = \varphi_\mathrm{l(r)}$ \cite{naik2017}. Consequently, the relative phase between the two coupling pathways $\varphi_c = \pi/2$ can be precisely set by controlling the relative phase between the flux modulation drives of the two couplers. We point out that, beyond shifting the emitter frequency by $\Delta$, the periodic flux modulation also creates additional undesired frequency components in the emitter's spectrum (separated by integer multiples of $\Delta$, hereafter referred to as the `sidebands' see \cite{Silveri2017Apr}), which can act as parasitic decay channels into the waveguide. To suppress these decay channels, each dissipation port in our experiment contains a compact microwave resonator, which filters the emission into the waveguide spectrally. \Cref{fig:1}b shows an image of the full device, with the emitter transmon, two frequency-tunable couplers, and the filter resonators. \Cref{fig:1}c shows the transmission spectrum through the waveguide for a weak coherent drive, where we can identify the resonant features corresponding to the filter resonators (R$_\mathrm{l,r}$), the couplers (C$_\mathrm{l,r}$), and the first-order sideband of the emitter qubit (E$_\text{1}$ at $\omega_\mathrm{E1}=\omega_\mathrm{E}+\Delta$). We deliberately design the filter resonators to have different resonance frequencies, with their detuning far exceeding their external decay rates to the waveguide. This condition is required to avoid mode hybridization between the resonators via the photon-mediated exchange interaction through the waveguide \cite{kockum2018a}. We also note that a similar concept with an alternative approach to sideband filtering has been theoretically proposed based on photonic crystal waveguides \cite{Wang2022May}. At optimal settings for chirality, we set the flux drive amplitudes of the couplers to achieve equal emitter-waveguide couplings via both dissipation ports (see \cref{fig:1}c). {A full analytical analysis of the parametric waveguide coupling and the spurious sideband suppression is provided in \cref{appendix:CMT_Full}.}

\subsection*{Device parameters}
In our experiment, the emitter is a transmon qubit with a maximum frequency of $\omega_\mathrm{E}/2\pi$ = 5.636 GHz. The tunable couplers are flux modulated with a frequency of $\Delta/2\pi = 805$ MHz, creating the emitter first (blue) sideband at a frequency of $\omega_\mathrm{E1}/2\pi = \omega_\mathrm{ge}/2\pi$ = 6.441 GHz (see \cref{fig:1}c). Flux control of the tunable couplers is enabled by SQUID loops with two symmetric Josephson junctions (see \cref{appendix:Methods}4). The tunable couplers are designed to operate in the transmon regime and are flux-biased to $\omega_\mathrm{C, l(r)}/2\pi$ = 6.482 (6.402) GHz. At the experiment operation settings, the filter cavity frequencies are $\omega_\mathrm{R, l(r)}/2\pi$ = 6.184 (6.712) GHz {(shifted from their `bare' values due to interaction with couplers)}. We control the external coupling rate $\Gamma^\mathrm{f}_\mathrm{1D}/2\pi$ of the chiral artificial atom by changing the couplers' configuration (see \cref{appendix:Methods}3). The distance between the two coupling points along the waveguide is 4.590 mm, which corresponds to a $\lambda/4$ separation at $\omega_\text{E1}/2\pi =$ 6.441 GHz ($\lambda ={c}/({f\sqrt{\varepsilon_\mathrm{eff}}})$, $\varepsilon_\mathrm{eff} = 6.45$). Our fabrication methods and device parameters are summarized in \cref{appendix:Methods}. 


\begin{figure*}[t!]
\centering
\includegraphics[width=1.02\textwidth]{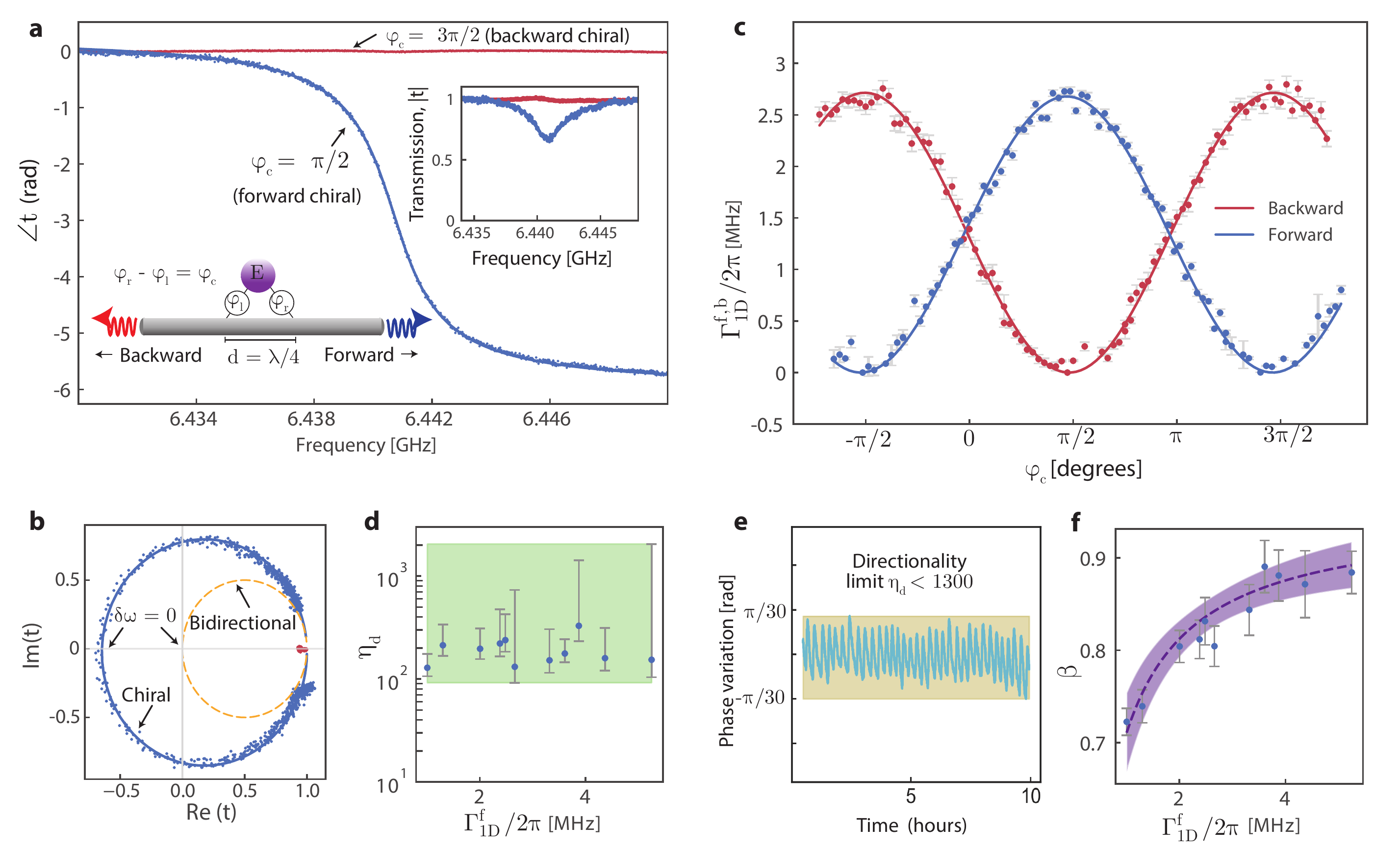}    
\caption{\textbf{Elastic chiral response under weak drives.} (a) The angle of the complex transmission coefficient measured with a weak microwave tone propagating in the forward direction through the waveguide. At $\varphi_\mathrm{c}$ = $\pi/2$ (blue curve), the atom is dominantly coupled to forward propagating modes ($\Gf \gg \Gb + \Gamma^{\prime}$) and we observe a $2\pi$ change in the phase imparted to the microwave tone as its detuning is swept about the resonance - a signature of chirality. When $\varphi_\mathrm{c} = 3\pi/2$ (red curve), the atom is chiral in the backward direction and the emitter interacts minimally with the forward propagating probe. The inset shows the amplitude of the transmission coefficient. Solid lines show fits to theory. (b) Measured transmission coefficient in the complex plane. At resonance ($\delta\omega = 0$), the transmission coefficient lies to the left of the origin in the case where the atom is dominantly coupled to forward propagating modes (blue curve). This results in the $2\pi$ phase change shown in (a). In contrast, in a non-directional system ($\Gf = \Gb$), the transmission coefficient would always stay to the right of the origin (yellow curve), and as a result, the transmission phase change cannot exceed $\pi$. (c) Periodic behavior of the emitter-waveguide coupling rate $\Gamma_\mathrm{1D}$ when the relative phase $\varphi_\mathrm{c}$ is varied. The blue (red) curve corresponds to the measurement with the probe propagating forward (backward). Solid lines are fits to theory. (d) Measured directionality ratio $\eta_d = \Gf/\Gb$ for various atom-waveguide coupling rates. (e) Measured fluctuations in the modulation phase of one of the couplers across 10 hours. The measurements correspond to a relative phase variance of $\langle \delta\phi_c^2\rangle = 3 \times 10^{-3}$ between the two couplers, and a phase-fluctuation limited bound of $\eta_d < 1.3\times 10^3$ on the directionality. (f) Measured values of the ratio between the atom's emission into the waveguide's mode of interest and its total decoherence rate, $\beta$ = $\Gf/(\Gf + \Gb + \Gamma^{\prime})$. The dotted purple line and shaded region show the fit results and the corresponding uncertainty for a model accounting for decoherence due to emitter-coupler hybridization and finite waveguide temperature.  All error bars denote 95\% confidence intervals.} 
    \label{fig:fig3}
\end{figure*}

\subsection*{Observation of chirality}
We characterize the response of the atom by performing transmission spectroscopy by applying a weak microwave drive of variable frequency to one end of the waveguide. This measurement is performed at a sufficiently low power such that the atom's excited state's population remains negligible. In a system with a partial directional response (the most general case), coherent scattering from the atom results in a Lorentzian lineshape with a transmission coefficient given by,
 \begin{align}
   t(\delta\omega) &= 1 - \dfrac{\Gf}{i\delta \omega + \Gamma_{\text{tot}}/2}, \nonumber \\
   \Gamma_{\text{tot}} & = \Gf + \Gb + \Gamma',     \label{eq:S21} 
\end{align}
where $\delta\omega = \omega - \omega_\mathrm{ge}$ is the atom-drive detuning. Here, $\Gamma_{\text{1D}}^\text{f,b}$ are the rates of the atom's spontaneous emission in the forward and backward directions, and $\Gamma'$ is its intrinsic decoherence rate. For a chiral atom in the strong coupling regime, when the atom couples dominantly to the forward propagating modes ($\Gf \gg \Gb + \Gamma^{\prime}$), we expect to see a $2\pi$ change in the phase imparted to the transmitted probe as the frequency is swept across the resonance. In contrast, this phase change cannot exceed $\pi$ when the atom-waveguide coupling to the forward and backward propagating modes are symmetric (see \cite{kono2018, besse2018,cohen2019} and \cref{appendix:Methods}2). 

\Cref{fig:fig3}a,b show the measurement results for two different phase settings. As evident, when $\varphi_\mathrm{c} = \pi/2$, we observe the canonical signature of chirality as a $2\pi$ phase across the resonance, which is consistent with our expectation for $\Gb/\Gf \rightarrow 0$. Conversely, when we set $\varphi_\mathrm{c} = 3\pi/2$ (via digital control of the phases of the couplers' flux drives), the atom's interaction with the forward-propagating drive disappears, consistent with $\Gf/\Gb \rightarrow 0$. This phase-sensitive directional behavior can be controlled with a fine resolution by varying $\varphi_\text{c}$ in small steps across a full $2\pi$ range.  \cref{fig:fig3}c shows the $\Gf$ obtained from a fit to the measured complex transmission coefficient, where we observe a sinusoidal dependence on the relative coupling phase. At each phase setting, we repeat the experiment with the drive tone propagating backward (this is done using a pair of electromechanical switches, see \cref{appendix:Methods}2) to obtain $\Gb$. As evident, the backward emission also varies periodically with  an out-of-phase profile with respect to the forward emission.

These observations verify our understanding of the underlying physical principles governing the operation of our device. We then proceed to benchmark directionality, defined as $\eta_d = \Gf/\Gb$ (at the optimal $\varphi_\text{c}$), as a figure of merit for a chiral atom. \cref{fig:fig3}d shows the extracted directionality bounds in a series of experiments in which we change the magnitude of atom-waveguide coupling by changing the emitter-coupler detuning. As evident, we can achieve $\eta_d \gtrsim 100$ consistently, which indicate a near-perfect chiral response. We note that the extracted bounds are conservative estimates limited by the sensitivity of our characterization technique (due to the difficulty of extracting a vanishingly small $\Gb$ from the spectral response, see \cref{appendix:Methods}2, \cref{appendix:chirality_bound}) and the actual directionality ratios are likely to be higher. As an additional test, we characterize the phase stability of the parametric drives, which can affect the chiral response due to the interferometric nature of our experiment. We do this by directly measuring the phase fluctuations of a coupler's modulation drive over a long time span and using it to calculate a bound on directionality analytically. We find that the phase fluctuations of the flux drives do not play a significant role in our experiments (see \cref{fig:fig3}e). 

Having established the chiral response, we now tend to a more general figure of merit in waveguide QED, namely the ratio between the atom's emission into the waveguide's mode of interest and its total decoherence rate, $\beta$ = $\Gf/(\Gf + \Gb + \Gamma^{\prime})$ \cite{lodahl2017}. The measured values of $\beta$ in our experiment are shown in \cref{fig:fig3}f. As the coupling to the backward propagating modes is suppressed nearly completely, the measured $\beta$ factors are dominantly determined by the intrinsic decoherence rate $\Gamma^{\prime}$. We observe an improvement in the measured $\beta$ values with increasing $\Gf$, consistent with a model assuming a constant $\Gamma^{\prime}$. From a fit to the data, we find $\Gamma^\prime/2\pi=  350 \pm 45$ kHz. The increasing trend of $\beta$-factors saturates at larger values of $\Gf$, indicating an increase in the decoherence rate concurrent with the increasing emission rate into the waveguide. This behavior may be attributed to increased decoherence due to the  finite temperature of the waveguide and an increase in the emitter-coupler mode hybridization (\cref{appendix:waveguide-temp}). The measured value of $\beta = 0.89 \pm 0.03$ in our experiment are comparable to the highest reported values in optical chiral systems \cite{sollner2015}. Moreover, we estimate that an order-of-magnitude improvement ($\Gf/\Gamma^{\prime} > 100$) is within reach in our system with improved device design (see \cref{appendix:waveguide-temp}), corresponding to $\beta$-factors close to unity.



\subsection*{Resonance fluorescence}
\begin{figure*}[t!]
\centering
\includegraphics[width=0.95\textwidth]{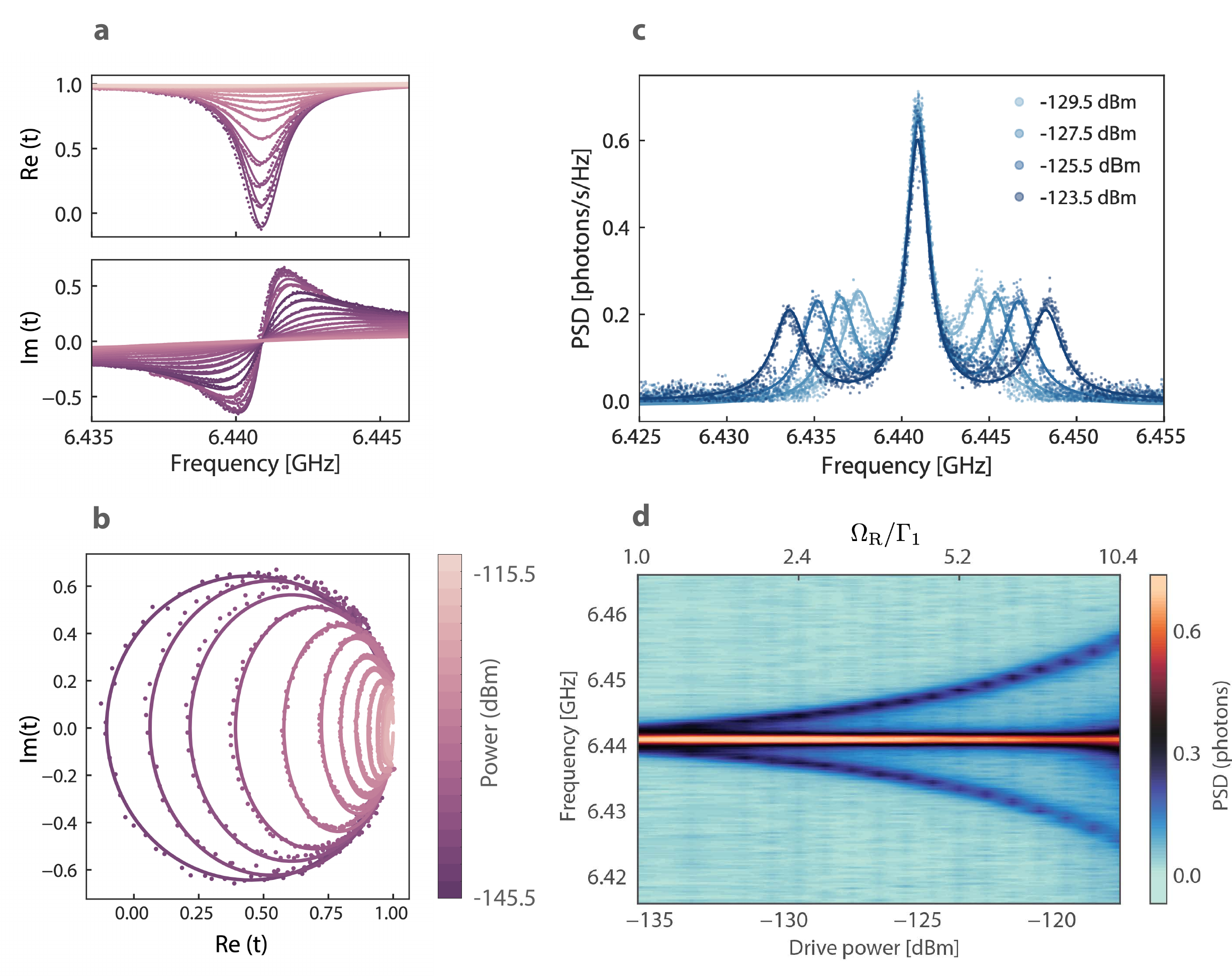}    
\caption{\textbf{Elastic and inelastic response of a strongly-driven chiral atom:} a) and b) Coherent response of the emitter with increasing drive power. The transmission amplitude $t$ is initially circular in the complex plane and becomes elliptical with increasing drive power, a signature of saturation of two-level atoms. Solid lines are fits to the expected theoretical response shown in  \cref{eq:power_broadening}. c) Measured resonance fluorescence spectrum of the chiral artificial atom, where we see well-resolved Mollow triplets. Solid lines are fits to \cref{eq:mollow_maintext}, showing good agreement between experiment and theory. We extract an energy relaxation rate $\Gamma_1/2\pi$ = 1.34 $\pm$ 0.15 MHz, decoherence rate $\Gamma_2/2\pi$ = 0.72 $\pm$ 0.08 MHz and a small pure dephasing rate of $\Gamma_{\phi}/2\pi = 50$ kHz for the chiral atom from these fits. d) Measured power spectral density (PSD) for a range of drive powers. The Rabi frequency $\Omega_\mathrm{R}$ is varied from a few hundred kHz to 14 MHz, indicating that the emitter can be driven strongly ($\Omega_\mathrm{R}/\Gamma_1 \approx$ 10).}
    \label{fig:fig4}
\end{figure*}
Our experiments so far have established a chiral response at low drive powers. In this regime, however, the elastic response of two-level systems is identical to that of bosonic systems such as cavities, which can also exhibit a unidirectional response \cite{petersen2014}. We next probe the quantum nonlinear behavior of the artificial atom, which manifests as the saturation of the elastic response. For a two-level system under strong drives, the transmission coefficient through the waveguide is given by (see \cref{appendix:mollow}), 
\begin{align}
t(\delta\omega)= 1 - \dfrac{\Gamma_\text{1D}^\mathrm{f}\Gamma_1 \left( \Gamma_2 - i\delta\omega \right)}{\Omega_\text{R}^2 \Gamma_2 + \Gamma_1(\delta\omega^2 + \Gamma_2^2)}.
\label{eq:power_broadening}
\end{align}
Here, $\Omega_\mathrm{R} = \sqrt{4P_\text{in}\Gamma_\mathrm{1D}^\mathrm{f}/\hbar\omega_\text{ge}}$ is the Rabi frequency from drive, $\Gamma_1$ is the total energy relaxation rate, $\Gamma_2 = \Gamma_1/2 + \Gamma_{\phi}$ is the total decoherence rate, and $\Gamma_{\phi}$ is the pure dephasing rate of the atom. \cref{fig:fig4}a show the results of a measurement in our systen, where increasing the incident power ($P_\text{in}$) leads to a monotonic reduction of the transmission phase. In the complex plane, this power-broadening effect manifests as a change in the shape of the trajectory from circular to elliptical (\cref{fig:fig4}b). The good agreement between the experiment and the fits indicates our model's validity, which assumes a single two-level atom with unidirectional coupling to a 1D bath. From these fits, we obtain $\Gamma_1/2\pi = 1.35 \pm 0.03$ MHz and $\Gamma_2/2\pi = 0.67 \pm 0.1$ MHz.

Beyond the elastic response, we also verify the two-level system behavior by measuring the resonance fluorescence spectrum. Under strong drives, when $\Omega_\mathrm{R} \gg \Gamma_\text{tot}$, inelastic scattering from an atom leads to the emergence of three distinct peaks in the emission spectrum (at $\omega_\mathrm{ge}$ and $\omega_\mathrm{ge} \pm \Omega_\mathrm{R}$) \cite{mollow1969, Astafiev2010Feb,lu2021, cottet}, known as the Mollow triplet. For a chiral atom, the power spectral density (PSD) of the incoherent emission is given by (see \cref{appendix:mollow}),
\begin{align}
\begin{split}\label{eq:mollow_maintext}
S(\omega) = \dfrac{1}{2\pi} & \dfrac{\hbar \omega_0 \Gamma_\mathrm{1D}^\mathrm{f}}{4} \left (\dfrac{\Gamma_\mathrm{s}}{\left(\delta\omega + \Omega_\mathrm{R}\right)^2 + \Gamma_\mathrm{s}^2}  \right. \\ & + 
\left. \dfrac{2\Gamma_2}{\delta\omega  + \Gamma_2^2} + 
\dfrac{\Gamma_\mathrm{s}}{\left(\delta\omega - \Omega_\mathrm{R}\right)^2 + \Gamma_\mathrm{s}^2} \right)
\end{split}
\end{align}

The measured resonance fluorescence spectra in our system are shown in \cref{fig:fig4}c and d, where we observe well-resolved Mollow triplets over a wide range of drive powers. Solid lines are fits to the model from \cref{eq:mollow_maintext}, showing good agreement to our model. From the fits, we extract $\Gamma_1/2\pi$ = 1.34 $\pm$ 0.15 MHz, $\Gamma_2/2\pi$ = 0.72 $\pm$ 0.08 MHz, in agreement with the rates extracted from the elastic response. The persistence of chirality under strong drives is clearly manifested by the constant total linewidth and the power under each peak, as well as the rate of scaling of the Rabi frequencies with the input power (see \cref{appendix:mollow}).
These measurements confirm that the chiral atom can be driven with large drive powers ($\Omega_\mathrm{R}/\Gamma_{1} \approx 10$). The upper limit on the drive power in our system is limited by the excitation of the couplers, which are weakly driven due to their small detuning to the emitter qubit. This constraint can be relaxed in future experiments by increasing the coupler-emitter detuning at the expense of an increased coupler-waveguide coupling.



\begin{figure*}[htbp]
\centering
\includegraphics[width=0.98\linewidth]{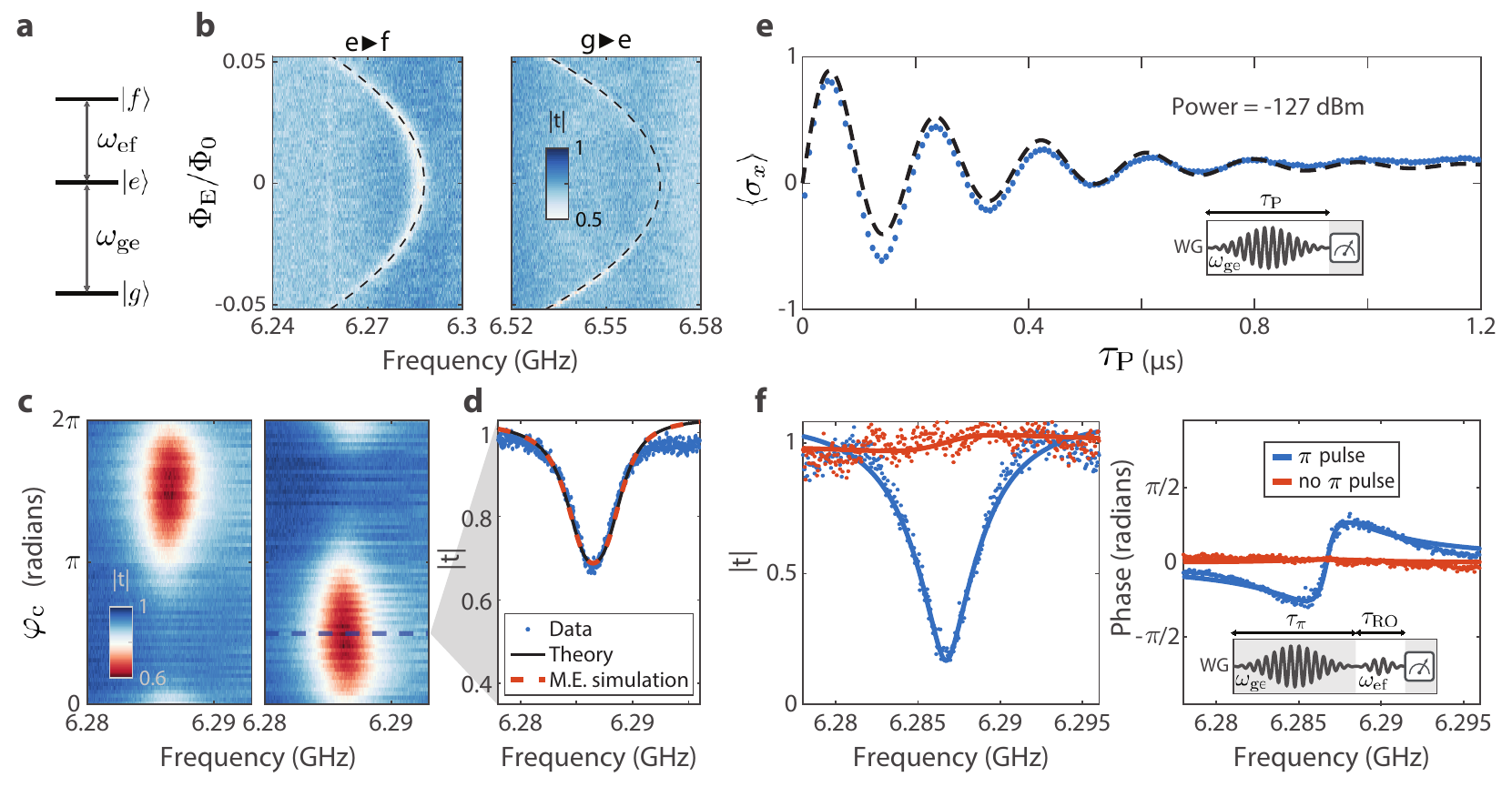}
\caption{\textbf{Chirality of the $\ef$ transition.} (a) Level structure of the emitter transmon. (b) Two-tone spectroscopy of emitter transmon. The $\gee$ transition is continuously driven on resonance, populating $|e\rangle$. A microwave probe tone then excites the $\ef$ transition. The emitter flux line is biased to tune the artificial atom frequency. (c) Transmission spectrum near $\ef$ transition as the relative phase between coupler drives ($\varphi_\mathrm{c}$) is varied. Left and right subpanels correspond to opposite excitation directions of the waveguide. We use $\Delta/2\pi = 930$ MHz for this measurement. (d) Transmission trace of chiral $\ef$ transition, with fitted theory and master equation simulation plots, overlaid. (e) Rabi oscillations for average drive power of -127 dBm. 
Inset: pulse sequence for Rabi oscillation measurements. The qubit is excited by a Gaussian pulse of variable length $\tau_\mathrm{P}$ via the waveguide. Qubit emission is averaged in a phase-coherent fashion to obtain $\langle \sigma_x \rangle$. (f) Transmission and phase of the $\ef$ transition under conditions for forward chirality, with (blue) and without (red) the application of the $\pi$ pulse. Inset: Measurement protocol for pulsed spectroscopy of the $\ef$ transition. The qubit is excited to $|e\rangle$ by a resonant $\pi$ pulse. A readout pulse then probes the $\ef$ transmission.}
\label{fig:4_e2f}
\end{figure*}

\subsection*{Chirality of the $\ef$ transition}
 Our experiments so far have established a chiral two-level system. However, the emitter transmon qubit in our experiment is a nonlinear oscillator with a rich level structure, including a third energy level $|f\rangle$ (see \cref{fig:4_e2f}a). The distinct $\ef$ transition frequency of a transmon has been used as an auxiliary degree of freedom for time-domain control of the emission and implementation of qubit-photon entangling gates \cite{kurpiers2018,VF2022,Kono2018Jun,reuer2022, Besse2018Apr}. Here, we demonstrate that similar functionalities can be implemented with our chiral qubit. 
 
 We begin by investigating the energy level structure of the chiral atom via two-tone spectroscopy, where we probe the system with a weak drive through the waveguide while populating the excited state $|e\rangle$ with a strong drive at $\omega_\mathrm{ge}$. The results are shown in \cref{fig:4_e2f}b, where we can identify a resonant feature at $\omega_\mathrm{ef}$. From this data, we obtain an anharmonicity of $\alpha = \omega_\mathrm{ge}-\omega_\mathrm{ef} = 283$ MHz, in close agreement with the designed anharmonicity value for the emitter qubit (272 MHz). We next find the chiral settings for the $e-f$ transition by repeating the two-tone spectroscopy while varying the relative coupling phase $\varphi_\mathrm{c}$.
 As described previously, achieving strong coupling to the waveguide requires cancellation of the backward emission rate by satisfying the $kd + \varphi_\mathrm{c} = \pi$ criterion. Additionally, we need to suppress parasitic sources of decay, which for the case of $e-f$ transition, includes the radiative decay of the $g-e$ transition into the waveguide. While the filter cavities in our device are designed for optimal spectral filtering near $\omega_\mathrm{ge}$, we can find a chiral operation point for the $e-f$ transition by properly choosing the modulation frequency $\Delta$. \cref{fig:4_e2f}c shows the results of transmission spectroscopy performed in the forward and backward directions, where we observe a periodic modulation in the contrast of the spectral feature at $\omega_\mathrm{ef}$ via $\varphi_\mathrm{c}$. At the optimal choice of $\varphi_\mathrm{c}$ extracted from this measurement (\cref{fig:4_e2f}d ), we confirm the chiral response for the $e-f$ transition using fits to a three-level system model and obtain a  directionality ratio $\eta_d$ of 12 (see \cref{appendix:EF}). 

We next employ time-domain control of the transmon to realize a conditional phase response. We begin with pulsed excitation of the emitter's $g-e$ transition through the waveguide. Due to the absence of a readout resonator in our experiment, we directly measure the emitted field from the qubit with heterodyne detection and phase-coherent averaging. This measurement provides the component of the qubit's Bloch vector in quadrature with the drive, $\exc{\sigma_x}$, which evolves with time as (see \cref{appendix:Rabi-oscillations}) 
\begin{align}
    \exc{\sigma_x}(t = \tp) = \sin(\Omega_\mathrm{R}\tp)\exp\left(- \Gamma_\mathrm{R}\tp\right).
    \label{eq:rabi-main-text}
\end{align}
Here, $\Gamma_\mathrm{R}$ = $(\Gamma_1 + \Gamma_2)/2$ and $\tau_\mathrm{P}$ is the duration of the driving pulse \cite{cottet}. \cref{fig:4_e2f}e shows the results, where we observe Rabi oscillations of the qubit as it is driven through and simultaneously decays into the open waveguide. Using these measurements, we calibrate a $\pi$-pulse to prepare the emitter in the excited state (\cref{fig:4_e2f}f inset). We then measure the complex transmission coefficient near the $e-f$ transition by sending a pulse centered at $\omega_\mathrm{ef}$ to the waveguide and performing heterodyne detection at the output. \cref{fig:4_e2f}f shows the measurement results, where we observe a phase change across the resonance when the qubit is initialized in the $|e\rangle$ state. Further, repeating the experiment without the $\pi$ pulse results in a flat spectral response near $\omega_\mathrm{ef}$. This state-dependent phase response can be combined with dispersion engineering (to protect the g-e transition from radiative decay to the waveguide) for future implementations of qubit-photon gates \cite{Kono2018Jun,reuer2022, Besse2018Apr} with chiral qubits (see \cref{appendix:EF}).

\subsection*{Conclusion}
In conclusion, we use distributed parametric interactions to break the time-reversal symmetry and realize unidirectional emission from a superconducting qubit into a planar microwave waveguide. In our system, we can achieve near-perfect directionality and strong coupling to the waveguide by controlling the forward and backward emission rates in situ and suppressing the parasitic sources of decay. We further verify the persistence of the two-level system behavior of the chiral qubit under strong drives ($\Omega_R/\Gamma_1 \approx 10$) with resonance fluorescence measurements. Finally, we show a directional response from the second transition of the chiral qubit and use it with pulsed control to realize a qubit-state-dependent phase response for traveling photonic wave packets. Our experiment thus provides an integrated platform for artificial chiral atoms with strong coupling to a one-dimensional photonic bath in the microwave domain. 
Looking ahead, we anticipate significant improvements in this platform by stronger suppression of parasitic decay using metamaterial waveguides  \cite{mirhosseini2018}, reducing dephasing with asymmetric junctions \cite{hutchings2017a}, and better waveguide thermalization with cryogenic attenuators \cite{yeh2017}. Implementing these measures is expected to lead to Purcell factors ($\Gf/\Gamma^{\prime}$) beyond 100 and further device miniaturization. With these improvements, we envision chip-scale experimental studies of chiral quantum optics with arrays of artificial atoms, which may enable quantum state transfer immune to thermal noise \cite{10.1103/physrevlett.118.133601,Xiang:2017jd}, quantum networks with all-to-all connectivity \cite{guimond2020,mahmoodian2016}, and driven-dissipative stabilization of many-body entanglement \cite{stannigel2012,mirza2016a}.

\section*{acknowledgments}
 This work was supported by startup funds from the Caltech EAS division, a Braun trust grant, and the National Science Foundation (grant No. 1733907). C.J. gratefully acknowledges support from the IQIM/AWS Postdoctoral Fellowship. F.Y. gratefully acknowledges support from the NSF Graduate Research Fellowship.

\clearpage

%

\appendix
\section{Methods}
\label{appendix:Methods}
\subsection{Fabrication}
Our device is fabricated on a 1 cm $\times$ 1 cm high-resistivity (10 k$\Omega$-cm) silicon substrate. Electron-beam lithography is used to pattern the structures in separate metal layers on the chip. Each lithography step is followed by electron-beam evaporation of metal and liftoff in N-methyl-2-pyrrolidone at 150$^\circ$ C for 1.5 hours. Device layers are as follows. (i) 150 nm thick niobium markers, deposited at 3\text{\normalfont\AA}/s. (ii) 120 nm thick aluminum ground plane, waveguide, flux lines, resonators, and qubit capacitors, deposited at 5\text{\normalfont\AA}/s. (iii) Josephson junctions evaporated (at 5\text{\normalfont\AA}/s) using double angle evaporation and consisting of 60 nm and 120 nm layers of aluminum, with 15 minutes of static  oxidation between layers. (iv) 150 nm thick aluminum band-aids and air-bridges, deposited at 5\text{\normalfont\AA}/s. Band-aids ensure electrical contact between Josephson junctions and qubit capacitors. Air-bridges are used to ensure the suppression of the slot-line modes in the waveguide \cite{Chen2014Feb}. Air-bridges are patterned using grey-scale electron-beam lithography and developed in a mixture of isopropyl alcohol and de-ionized water, followed by 2 hours of reflow at 105$^\circ$ C \cite{Painter2020Dec}. Electron beam evaporation of the band-aid/bridge layer is preceded by 7 minutes of Ar ion milling.

\begin{figure}[htbp]
\centering
\includegraphics[width=1\linewidth]{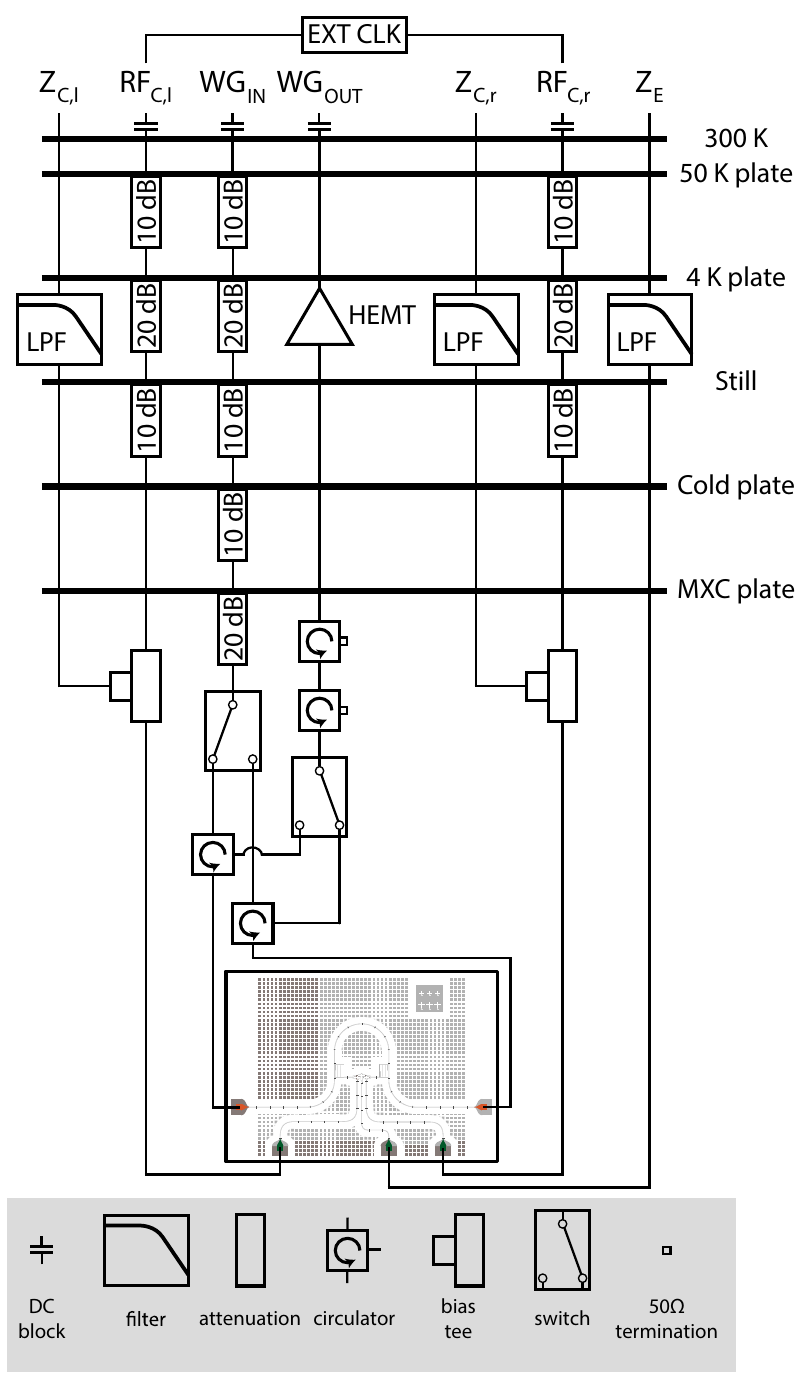}
\caption{\textbf{Measurement Setup} schematic of dilution fridge wiring for measurement.}
\label{fig:Fridge}
\end{figure}

\subsection{Measurement setup}
Measurements are performed in a $^3$He/$^4$He dilution refrigerator. A schematic of our measurement setup is shown in \cref{fig:Fridge}. The fabricated chip is wire-bonded to a PCB and placed in an aluminum box. The box is then mounted to the mixing plate which is cooled to a base temperature of 10 mK. 

The waveguide input line ($\mathrm{WG_{\mathrm{IN}}}$) is used to excite the chiral qubit from either waveguide direction. Toggling both switches in \cref{fig:Fridge} changes the excitation direction for transmission measurements. The waveguide input line ($\mathrm{WG_{\mathrm{IN}}}$) is attenuated at each temperature stage to minimize thermal noise; the total attenuation is 70 dB. A tunable attenuator (not shown) is also added to the input line at room temperature to control input power. Two isolators are used to reduce thermal noise in the waveguide output line ($\mathrm{WG_{\mathrm{OUT}}}$). The output is amplified by a high electron mobility transistor (HEMT) amplifier at the 4 K stage and a room temperature amplifier (not shown) outside of the fridge.

A low noise, multi-channel DC source provides current biases ($\mathrm{Z_{\mathrm{C,l}}}$, $\mathrm{Z_{\mathrm{C,r}}}$, $\mathrm{Z_{\mathrm{E}}}$) to flux tune the emitter and coupler qubit frequencies. A low-pass RF filter (32 kHz cutoff frequency) suppresses high-frequency thermal noise in the DC lines, which are not attenuated. Coupler DC bias lines are combined with RF inputs ($\mathrm{RF_{\mathrm{C,l}}}$, $\mathrm{RF_{\mathrm{C,r}}}$) using microwave bias tees, enabling frequency tuning and parametric modulation of coupler frequencies. Coupler RF drives (Rohde and Schwarz SMB 100A) are clocked by an external Rubidium clock, allowing for relative phase tuning. RF inputs are attenuated at the 50K, 4K, and cold plates to reduce thermal noise. 

\begin{figure}[htbp]
\centering
\includegraphics[width=0.9\linewidth]{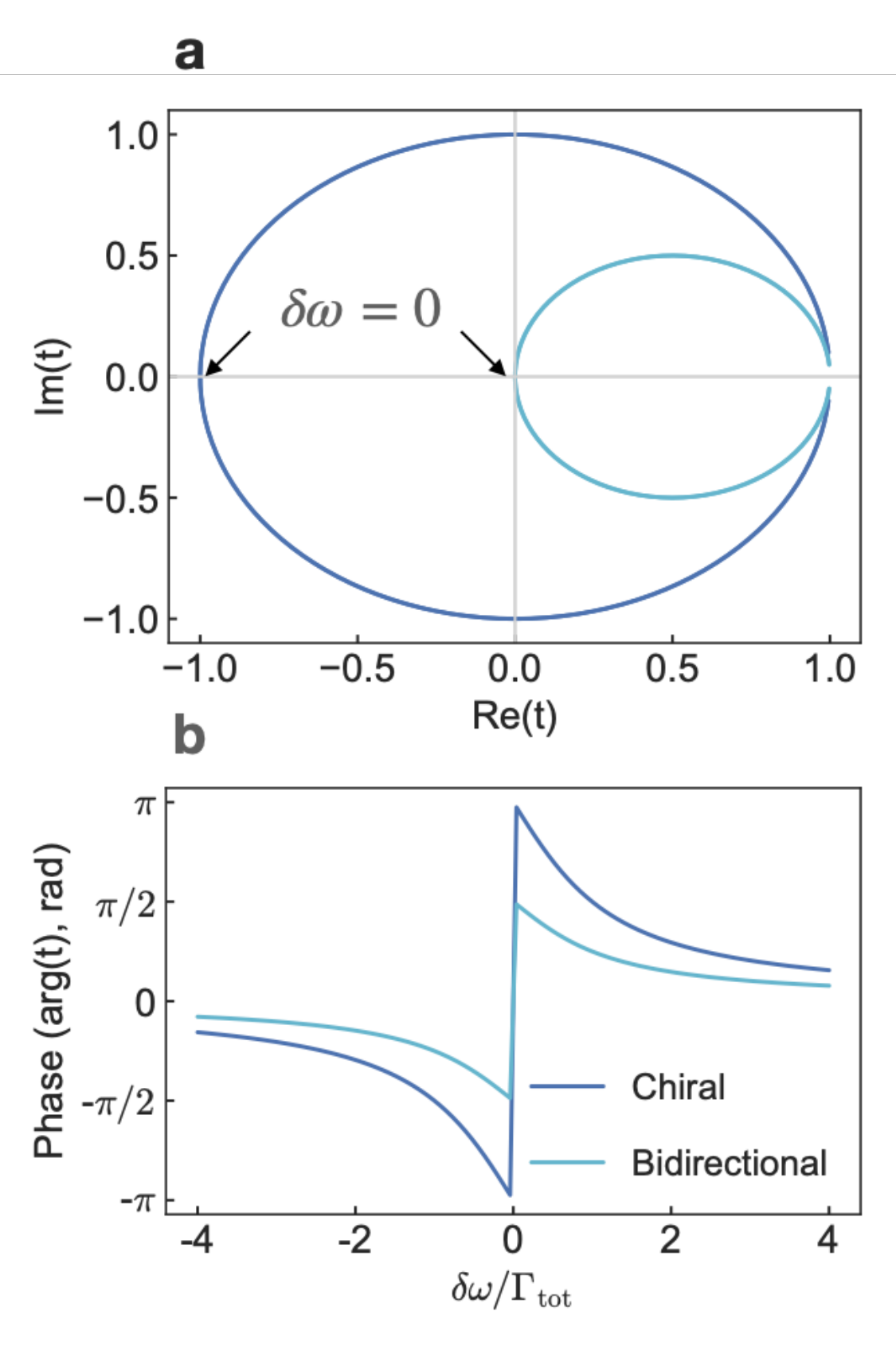}
\caption{(a) Complex transmission coefficient $t$ calculated from \cref{eq:input-out-single-sided} when the atom-waveguide coupling is chiral (dark blue) and bidirectional (light blue). Here, we assume that the intrinsic decoherence rate $\Gamma^\prime$ is small compared to coupling to waveguide modes. In the chiral case, symmetry of the atom-waveguide coupling to forward and backward propagating modes is broken and the atom can dominantly couple to forward propagating modes such that ($\Gamma_\mathrm{1D}^\mathrm{f}/(\Gamma_\mathrm{1D}^\mathrm{b} + \Gamma^\prime) >1$). At resonance ($\delta\omega = 0$), the transmission coefficient lies to the left of the origin $(\mathrm{Re}(t) < 0)$, and is equal to -1 when $\Gamma^\prime = 0$. In this case, there is a full $2\pi$ change in the phase imparted to the transmitted field as the detuning $\delta\omega$ is swept around the resonance (see (b)). When the atom-waveguide coupling is symmetric for the forward and backward propagating modes ($\Gamma_\mathrm{1D}^\mathrm{f} = \Gamma_\mathrm{1D}^\mathrm{b}, \Gamma_\mathrm{1D}^\mathrm{f}/(\Gamma_\mathrm{1D}^\mathrm{b} + \Gamma^\prime) \leq 1$), the transmission coefficient lies to the right of the origin ($\mathrm{Re}(t) \geq 0$) and is equal to zero when $\Gamma^\prime = 0$. In this case, the total change in the phase imparted to the transmitted field cannot exceed $\pi$ (see (b)).}
\label{fig:phase}
\end{figure}

\paragraph*{Transmission measurements}
The coherent response of the device is measured using a vector network analyzer (VNA, Agilent N5242A). The VNA can be used to simultaneously measure the amplitude and phase of the transmitted signal via heterodyne detection. For the characterization of emitter qubit chirality, the drive from the VNA is attenuated to sub-single-photon power levels in order to ensure that the qubit is not saturated. To obtain the atom-waveguide coupling rate $\Gamma_\text{1D}^\mathrm{f}$ and the total linewidth $\Gamma_\mathrm{tot}$, we use the circle-fit method when the condition $\Gamma_\text{1D}^\mathrm{f} > \Gamma_\text{1D}^\mathrm{b} + \Gamma^{\prime}$ is satisfied. The circle-fit method does not rely on initial conditions and provides more robust estimation of the coupling rates \cite{probst2015}. When $\Gamma_\text{1D}^\mathrm{f} \le \Gamma_\mathrm{1D}^\mathrm{b} + \Gamma^{\prime}$, we use nonlinear-least squares fitting to \cref{eq:input-out-single-sided} to obtain $\Gamma_\text{1D}^\mathrm{f}$ and $\Gamma_\mathrm{tot}$. 
When the atom-waveguide coupling is chiral and the atom couples dominantly to forward propagating modes ($\Gamma_\mathrm{1D}^\mathrm{f} > \Gamma_\mathrm{1D}^\mathrm{b} + \Gamma^{\prime}$), the transmission coefficient at resonance ($\delta\omega = 0$) lies to the left of the origin in the complex plane (see \cref{fig:phase}). This results in a full $2\pi$ change in the phase imparted to the transmitted field. In contrast, when the atom-waveguide coupling to forward and backward propagating modes is symmetric ($\Gamma_\mathrm{1D}^\mathrm{f} = \Gamma_\mathrm{1D}^\mathrm{b}, \Gamma_\mathrm{1D}^\mathrm{f} \leq \Gamma_\mathrm{1D}^\mathrm{b} + \Gamma^{\prime}$), the transmission coefficient lies to the right of the origin at resonance and the phase change imparted to the transmitted field cannot exceed $\pi$ (\cref{fig:phase}). 
Our current measurement setup \cref{fig:Fridge} is designed to perform transmission measurements in both forward and backward directions through the waveguide. 


\paragraph*{Resonance fluorescence measurements}
Resonance fluorescence measurements are performed with an RF spectrum analyzer (SA, Rohde and Schwarz FSV3013); the microwave excitation tone is provided by the VNA in zero-span mode. The spectrum analyzer acquisition is performed with a resolution bandwidth of 20 kHz. The background power level in this measurement is determined by the HEMT noise temperature T$_\text{HEMT}$ of 3.5 K. Using Bose-Einstein statistics, this corresponds to a power spectral density (PSD) of 11 photons/s/Hz at the emitter frequency ($\omega_\mathrm{ge} = 6.441$ GHz).  The HEMT background is subtracted from the signal traces, and the resulting power is normalized with the resolution bandwidth and the gain of the output line to obtain the PSD shown in \cref{fig:fig4}a and b. The output line gain is calibrated using thermometry measurements \cite{joshi2022}. We note that data shown in \cref{fig:fig4}a and b corresponds to PSD in linear frequency and is equal to $2\pi \times S(\omega)$, where $S(\omega)$ is given by \cref{eq:mollow_maintext}. 

\paragraph*{Time-domain measurements}
Time-domain measurements and pulsed excitations of the device are performed using the Quantum Machines OPX+ (QM) module, which is capable of arbitrary waveform generation and  heterodyne detection. To generate the drive, MHz frequency IF signals from the QM module are upconverted via mixing with a local oscillator (LO) supplied by an RF signal generator (Rohde and Schwarz SMB100) using IQ mixers (Marki Microwave MMIQ-0520LS). For readout, the signal from the output line is downconverted using an IQ mixer, and the resulting IF-frequency signal is demodulated. The duration of the $\pi$-pulse for the $\gee$ transition is determined from measurements of qubit Rabi oscillations. For the Rabi oscillation curves, the output is averaged in a phase-sensitive manner to obtain the qubit emission in quadrature with the drive (see also \cref{appendix:Rabi-oscillations}). For spectroscopy of the $\ef$ transition, the excited state $|e\rangle$ is first fully populated by driving $\gee$ with the calibrated $\pi$-pulse. Spectroscopy is then performed by probing the $\ef$ transition with a resonant readout pulse. Averaging is performed over $4\times10^6$ samples to obtain the results shown in \cref{fig:4_e2f}e. 


\begin{figure}[htbp]
\centering
\includegraphics[width=0.95\linewidth]{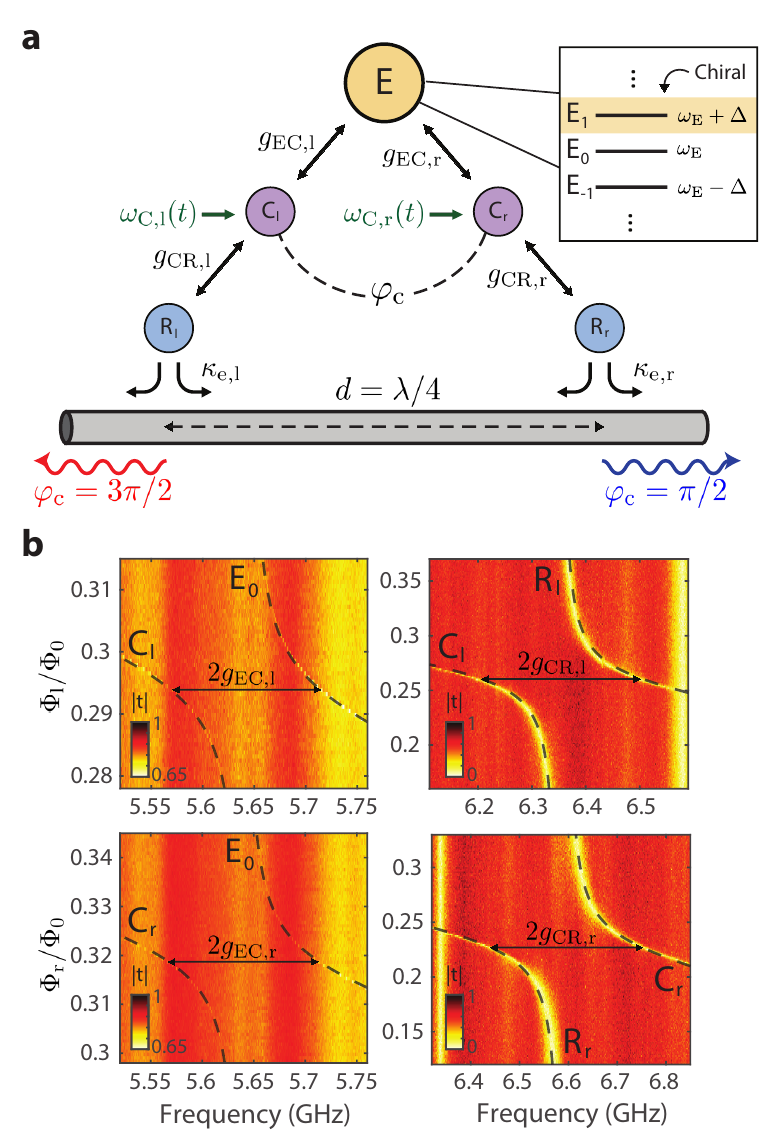}
\caption{\textbf{Atom-waveguide coupling scheme} (a) Schematic of the chiral atom-waveguide coupling scheme. The emitter couples to the waveguide at two points, separated by $d = \lambda/4$. In each decay pathway, a flux controlled coupler (purple, C$_{\mathrm{r,l}}$) is capacitively coupled to the emitter with a coupling strength $g_\mathrm{EC}$ and to a filter cavity (blue, R$_\mathrm{r,l}$) with a coupling strength $g_\mathrm{CR}$. The filter cavities are capacitively coupled to the waveguide with decay rates $\kappa_\mathrm{e, r (l)}$. The flux bias applied to the right (left) tunable coupler is sinusoidally modulated with an RF frequency $\Delta$, resulting in time-varying coupler frequencies $\omega_\mathrm{C, r(l)}(t) = \omega_\mathrm{C, r(l)} + \epsilon_\mathrm{l(r)}\sin(\Delta t + \varphi_\mathrm{r (l)}$). This creates an effective time-dependent emitter-waveguide coupling with a relative phase $\varphi_c = \varphi_r - \varphi_l$ between the two decay pathways. Setting $\varphi_\mathrm{c} = \pi/2$ results in forward chirality (blue), while $\varphi_\mathrm{c} = 3\pi/2$ results in backward chirality (red). (b) Avoided crossings of couplers with the emitter and filter resonators, obtained by flux tuning the frequency of the couplers. The top (bottom) row corresponds to the left (right) coupler, and the left (right) column corresponds to emitter-coupler (coupler-resonator) avoided crossings.}
\label{fig:coupler}
\end{figure}

\subsection{Emitter-waveguide interaction using tunable couplers}

The frequency-selective dissipation ports mediate the emitter decay to each waveguide coupling point. Each port contains a flux-controlled tunable coupler and a filter cavity. \cref{fig:coupler}a gives a schematic for the full device. Each coupler (purple, C$_{\mathrm{r,l}}$) is capacitively coupled to the emitter with a coupling strength $g_\mathrm{EC}$ and to the filter cavity (blue, R$_\mathrm{r,l}$) with coupling strength $g_\mathrm{CR}$. The filter cavities are realized as compact resonators with capacitive coupling to the waveguide. The emitter-coupler and cavity-coupler coupling rates are shown in \cref{fig:coupler}b and \cref{tab:table3} and are extracted by flux tuning the couplers into avoided crossings with the emitter and resonators.

\subsection{Device parameters}

\begin{figure*}[htbp]
\centering
\includegraphics[width=0.9\linewidth]{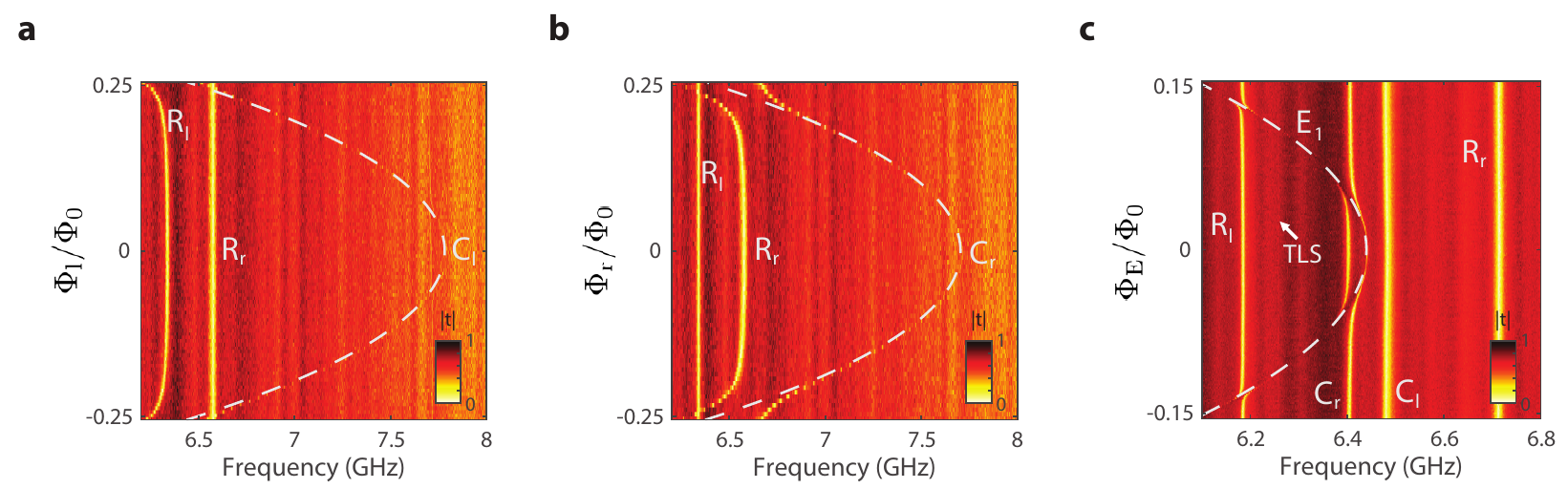}
\caption{\textbf{Flux tuning curves.} (a) Left and (b) right coupler flux tuning curves, obtained from VNA measurements. No crosstalk corrections are made here, and fits are used to extract mutual inductances ($M_\mathrm{LL}$ and $M_\mathrm{RR}$). (c) The tuning curve for the emitter's first blue sideband. The settings here correspond to the chiral configuration of Figure \ref{fig:1}c (bottom panel). An RF modulation of frequency $\Delta = $ 805MHz is applied to the couplers. A two-level system (TLS) defect coupled to the system is marked with the arrow.}
\label{fig:tuning_curves}
\end{figure*}

The emitter and couplers in our device are superconducting qubits designed in the transmon regime to mitigate charge noise. Bare qubit parameters are provided in Table \ref{tab:table1}. Max frequency ($\omega_\mathrm{max}/2\pi$) and Josephson energy ($E_\mathrm{J}$) are extracted from fits to DC flux tuning curves, shown in \cref{fig:tuning_curves}. Charging energy ($E_\mathrm{C}$) is obtained by measuring qubit anharmonicity via two-tone spectroscopy. A single microwave tone drives the qubit $|g\rangle\xrightarrow{}|e\rangle$ transition, which populates the $|e\rangle$ state. The $|e\rangle\xrightarrow{}|f\rangle$ transition then becomes visible under waveguide spectroscopy.

\begin{table}[htpb]
\begin{ruledtabular}
\begin{tabular}{ccccc}
\textrm{Qubit}&
\textrm{$\omega_\mathrm{max}/2\pi$ [GHz]}&
\textrm{$E_\mathrm{J}$ [GHz] }&
\textrm{$E_\mathrm{C}$ ($-\alpha$) [MHz]}&
\textrm{$E_\mathrm{J}/E_\mathrm{C}$}\\
\colrule
E               & 5.636 & 15.47 & 283.0 & 54.66 \\
C$_\mathrm{l}$  & 7.779 & 25.35 & 324.0 & 78.24 \\
C$_\mathrm{r}$  & 7.699 & 25.63 & 313.5 & 81.76 \\
\end{tabular}
\end{ruledtabular}
\caption{\label{tab:table1}
The Parameters for the emitter and coupler transmons.
}
\end{table}

The compact resonators consist of an inductive meander and capacitive `claw'. The latter section is shaped to engineer the coupler-resonator and resonator-waveguide couplings. Because detuning between resonators exceeds their individual external coupling rates, the waveguide-mediated exchange interaction and correlated decay can be safely neglected \cite{kockum2018a}. The resonator parameters are provided in Table \ref{tab:table2}. 

\begin{table}[htpb]
\begin{ruledtabular}
\begin{tabular}{cccc}

\textrm{Cavity} &
\textrm{$\omega/2\pi$ [GHz]} &
\textrm{$\kappa_\mathrm{e}$ [MHz]} &
\textrm{$\kappa_\mathrm{i}$ [MHz]} \\
\colrule
R$_\mathrm{l}$ & 6.337 & 24.95 & 0.451 \\
R$_\mathrm{r}$ & 6.577 & 41.74 & 0.187 \\
\end{tabular}
\end{ruledtabular}
\caption{\label{tab:table2}
Filter cavity parameters.
}
\end{table}

Capacitive couplings in our device can be expressed in terms of frequencies, mutual capacitances, self-capacitances, and self-inductances of the two relevant modes. Approximating qubits as linear oscillators and assuming only nearest-neighbor coupling, the interaction strength $g_{ij}$ is given below.
\begin{equation}
g_{ij} = \frac{1}{2}\frac{C_\mathrm{m}}{\sqrt{(C_i+C_\mathrm{m})(C_j+C_\mathrm{m})}}\sqrt{\omega_i \omega_j} \\
\end{equation}
Here, $C_\mathrm{m}$ is the mutual capacitance, $C_{i,j}$ are bare self-capacitances, and $\omega_{i,j}$ are adjusted frequencies of each mode. For two linear coupled oscillators, $\omega_{i,j} = \sqrt{\frac{C_{j,i}+C_\mathrm{m}}{L_{i,j}C^2_\Sigma}}$, where $C^2_\Sigma = C_i C_j + C_i C_\mathrm{m} + C_j C_\mathrm{m}$. For transmon qubits approximated as linear oscillators, $L = (\frac{\hbar}{2e})^2\frac{1}{E_\mathrm{J}}$, where $E_\mathrm{J}$ is the Josephson energy. Capacitive couplings in our device are extracted from avoided crossings observed in waveguide spectroscopy (\cref{fig:coupler} and are listed in Table \ref{tab:table3}.

\begin{table}[htpb]
\begin{ruledtabular}
\begin{tabular}{cc}

\textrm{Coupling} &
\textrm{Value [MHz]}  \\
\colrule
$g_\mathrm{EC,l}$ (Emitter, Left Coupler) & 72.65 \\
$g_\mathrm{EC,r}$ (Emitter, Right Coupler) & 73.15 \\
$g_\mathrm{CR,l}$ (Left Coupler, Left Resonator) & 149.50 \\
$g_\mathrm{CR,r}$ (Right Coupler, Right Resonator) & 155.55 \\
\end{tabular}
\end{ruledtabular}
\caption{\label{tab:table3}
Coupling Strengths.
}
\end{table}

\subsection{Flux biasing crosstalk}

The emitter and coupler frequencies are tuned via flux lines (Z-lines), which are biased with a low-noise, multi-channel DC source. The current applied to Z-line control the magnetic flux through a SQUID loop. Because of close proximity, the Z-lines experience cross-talk. In order to achieve independent frequency tuning in presence of this cross-talk, we extract a cross-inductance matrix for the system of the emitter and couplers. 

\begin{equation}
\begin{bmatrix} \Phi_\mathrm{L} \\ \Phi_\mathrm{E} \\ \Phi_\mathrm{R} \end{bmatrix} = 
\begin{bmatrix} 
M_\mathrm{LL} & M_\mathrm{LE} & M_\mathrm{LR} \\ M_\mathrm{EL} & M_\mathrm{EE} & M_\mathrm{ER} \\ M_\mathrm{RL} & M_\mathrm{RE} & M_\mathrm{RR}
\end{bmatrix}
\begin{bmatrix} I_\mathrm{L} \\ I_\mathrm{E} \\ I_\mathrm{R} \end{bmatrix}
\end{equation}

Here, L, E, and R represent the left coupler, emitter, and right coupler, respectively. The matrix elements are extracted by fits to the corresponding tuning curve. Two of the elements $M_\mathrm{EL}$ and $M_\mathrm{ER}$ are not measured due to the absence of direct coupling between the emitter and the waveguide. To correct for the small effect of $M_\mathrm{EL}$ and $M_\mathrm{ER}$, the emitter qubit is biased to the flux sweet spot prior to all measurements. The extracted inductance matrix is given below.

\begin{equation}
\textbf{M} = 
\begin{bmatrix} 
1.382 & -0.058 &  -0.065 \\ 0 & 1.086 & 0 \\ 0.095 & 0.052 & 1.398
\end{bmatrix}
\text{pH}
\end{equation}

\section{Input-output relations using SLH formalism}
\label{appendix:input-output}
We derive the input-output relations using the SLH formalism \cite{combes2017}. The SLH formalism is a powerful toolbox for modeling cascaded quantum systems and is naturally suited for obtaining master equations and input-output relations for giant atoms \cite{kockum2018a}. We model our chiral emitter as a giant atom with two coupling points, with a relative coupling phase $\varphi_\mathrm{c} = \varphi_\mathrm{r} - \varphi_\mathrm{l}$ between the two feet of the giant atom (as shown in \cref{fig:1}a). As shown in \cref{eq:atom-waveguide-hamiltonian}, the field of emission at each coupling point acquires a well-defined phase $\varphi_\mathrm{l(r)} = \mathrm{arg}[\tilde{g}_\mathrm{l(r)}]$ due to the complex coupling mediated by the time-modulated couplers between the atom and the waveguide.
In the rotating frame of the drive, we decompose the ($\mathbf{S},\mathbf{L},H$) triplets for the two coupling points as follows, 
\begin{align}
    G_\mathrm{f,l} & =  (1, \sqrt{\dfrac{\kappa_\mathrm{em}}{2}}\hat{\sigma}_{-}, \dfrac{\delta\omega}{2}\hat{\sigma}_{z}) \\
    G_\mathrm{f,r} & = (1, \sqrt{\dfrac{\kappa_\mathrm{em}}{2}}e^{i\varphi_\mathrm{c}}\mathrm{\hat{\sigma}}_{-}, 0) \\
    G_\mathrm{b,l} & =  (1, \sqrt{\dfrac{\kappa_\mathrm{em}}{2}}\mathrm{\hat{\sigma}}_{-}, 0) \\
     G_\mathrm{b,r} & =  (1, \sqrt{\dfrac{\kappa_\mathrm{em}}{2}} e^{i\varphi_\mathrm{c}}\hat{\sigma}_{-}, 0) 
\end{align}
where $\delta\omega = \omega_\mathrm{ge} - \omega$ is the detuning of emitter qubit at $\omega_\mathrm{ge}$ from the drive at frequency $\omega$, $\kappa_\text{em}$ is the magnitude of the decay rate of the giant atom at each of the two feet, f and b denote forward and backward propagating modes, respectively and $\mathrm{l(r)}$ denote the left and right coupling points. We assume that the system is driven by a forward-propagating coherent tone with complex amplitude $\alpha$ and write the ($\mathbf{S}, \mathbf{L}, H$) triplets for the drive ($G_\mathrm{f, drive}$) and the waveguide propagation ($G_\mathrm{WG}$) as, 
\begin{align}
    G_\mathrm{f, drive} = (1, \alpha, 0) \\
    G_\mathrm{WG} = (e^{i\varphi_\mathrm{WG}}, 0, 0)
\end{align}
where $\varphi_\mathrm{WG} = \omega_\mathrm{ge}d/v$ is the accumulated propagation phase between the two coupling points (see \cref{fig:1}a). The SLH triplet for the forward and backward moving parts is then given by, \begin{widetext}
\begin{align}
    G_\mathrm{f} & =  G_\mathrm{f,r} \triangleleft G_\mathrm{WG} \triangleleft G_\mathrm{f,l} \triangleleft G_\mathrm{f, drive}, \\
    G_\mathrm{b} & = G_\mathrm{b,l} \triangleleft G_\mathrm{WG} \triangleleft G_\mathrm{b,r}.
    \end{align}
Using cascading rules from the SLH formalism, we obtain the following SLH triplets for the system $G_\text{tot} = G_\mathrm{f} \boxplus G_\mathrm{b}$ and obtain \cite{kockum2018a}, 

\begin{align}
S_\text{tot} & = 
\begin{pmatrix}
    e^{i\varphi_\mathrm{WG}} & 0\\
    0 & e^{i\varphi_\mathrm{WG}} 
\end{pmatrix} \\
L_\text{tot} & = 
\begin{pmatrix}
    \alpha e^{i\varphi_\mathrm{WG}} + \sqrt{\dfrac{\kappa_\text{em}}{2}} (e^{i\varphi_\mathrm{WG}} + e^{i\varphi_\mathrm{c}})\hat{\sigma}_{-}\\
    \ke (1 + e^{i(\varphi_\mathrm{WG} + \varphi_\mathrm{c})})\hat{\sigma}_{-}
\end{pmatrix} \\
\dfrac{H_\text{tot}}{\hbar} & = \dfrac{\delta\omega}{2} \hat{\sigma}_z 
  -i \ke[\alpha \hat{\sigma}_{+} (1 + e^{i(\varphi_\mathrm{WG} - \varphi_\mathrm{c})}) -\alpha^\ast \hat{\sigma}_{-} (1 + e^{-i(\varphi_\mathrm{WG} - \varphi_\mathrm{c})})]. \label{eq:Hamiltonian}
\end{align}
\end{widetext}
Here, $S_\mathrm{tot}$ is the scattering matrix of the system, $L_\mathrm{tot} = (L_\mathrm{f}, L_\mathrm{b})^\top$ denotes the collapse operator for the forward $(L_\mathrm{f})$ and backward ($L_\mathrm{b}$) propagating modes and $H_\mathrm{tot}$ is the system Hamiltonian. The input-output relations can be written in terms of the collapse operators for the forward and backward propagating modes, 
\begin{align}
t = \dfrac{\langle L_\mathrm{f}\rangle}{\alpha_\mathrm{in}} = e^{i\varphi_\mathrm{WG}} + \dfrac{1}{\alpha_\mathrm{in}} \ke (e^{i\varphi_\mathrm{c}} + e^{i\varphi_\mathrm{WG}}) \langle \hat{\sigma}_{-} \rangle\nonumber \\
r = \dfrac{\langle L_\mathrm{b} \rangle}{\alpha_\mathrm{in}} =  \dfrac{1}{\alpha_\mathrm{in}} \ke (e^{i(\varphi_\mathrm{c} + \varphi_\mathrm{WG})} + 1)\langle \hat{\sigma}_{-} \rangle
\label{eq:SLH-input-output}
\end{align}
where $t$ and $r$ are the complex transmission and reflection coefficients, respectively. From \cref{eq:SLH-input-output} it is clear that the emission in the backward direction can be nulled when the condition is $\varphi_\mathrm{WG} + \varphi_\mathrm{c} = \pi$ is satisfied. This is the interference condition to obtain perfect chiral behavior. \cref{eq:SLH-input-output} can be further simplified by calculating $\langle \hat{\sigma}_{-} \rangle$ using the Heisenberg equation of motion $\langle  {\dot{\hat{\sigma}}_{-}} \rangle = -i[\hat{\sigma}_{-}, H_\text{tot}] - (\Gamma_\text{tot}/2) \hat{\sigma}_{-}$ = 0.  Here, $\Gamma_\text{tot}$ is the total decay rate of the emitter, including radiative and non-radiative decay, and dephasing. In the limit of weak drive such that the two-level system is not saturated, we obtain in steady state, 
\begin{align}
    \langle \hat{\sigma}_{-} \rangle = \dfrac{-\ke \alpha [1 + e^{i({\varphi_\mathrm{WG} - \varphi_\mathrm{c}})}]}{i\delta\omega + \dfrac{\Gamma_\text{tot}}{2}}
    \label{eq:sigma_minus}
\end{align}
Combining \cref{eq:SLH-input-output} and \cref{eq:sigma_minus}, we obtain the transmission coefficient
\begin{align}
    t = e^{i\varphi_\mathrm{WG}} \left( 1- \dfrac{\kappa_\text{em}[1 + \cos{(\varphi_\mathrm{c} - \varphi_\mathrm{WG}})]}{i\delta\omega + \dfrac{\Gamma_\text{tot}}{2}}\right)
    \label{eq:s21_SLH}
\end{align}
From \cref{eq:s21_SLH}, we identify the effective atom-waveguide coupling for the forward propagating modes, 
\begin{align}
\Gamma_\text{1D}^\mathrm{f} = \kappa_\text{em}[1 + \cos{(\varphi_\mathrm{c} - \varphi_\mathrm{WG})}]    
\label{eq:forward_gamma1D}
\end{align}
The transmission can be written in terms of $\Gamma_\mathrm{1D}^\mathrm{f}$ to be (up to a global phase factor), 
\begin{align}
    t = e^{i\varphi_\mathrm{WG}} \left(1 - \dfrac{\Gamma_\text{1D}^\mathrm{f}}{i\delta\omega + \dfrac{\Gamma_\text{tot}}{2}}\right)
    \label{eq:input-out-single-sided}
\end{align}
Note that the transmission expression is equivalent to that of a single-sided cavity with an external coupling rate $\Gamma_\text{1D}^\mathrm{f}$. The emitter-waveguide coupling for the backward propagating modes can be similarly evaluated by assuming a drive from the right side of the waveguide and repeating the analysis above. The coupling rate for the backward propagating modes is then given by
\begin{align}
\Gamma_\text{1D}^\mathrm{b} = \kappa_{\text{em}}[1 + \cos{(\varphi_\mathrm{c} + \varphi_\mathrm{WG})}]   
\label{eq:backward_gamma1D}
\end{align}

We write the master equation for the case where the condition for perfect chirality is satisfied. Simplifying \cref{eq:Hamiltonian} for the case $\varphi_\mathrm{WG} = \pi/2, \varphi_\mathrm{c} = \pi/2$, we obtain the Hamiltonian
\begin{align}
\dfrac{H_\text{tot}}{\hbar} = \dfrac{\delta\omega}{2}\hat{\sigma}_z + \frac{1}{i} \sqrt{\dfrac{{\kappa_\text{em}}}{2}}[2\alpha \hat{\sigma}_{+} - 2\alpha^{\ast}\hat{\sigma}_{-}] 
\end{align}
Assuming without loss of generality that the drive $\alpha$ is real, and using the relations $\hat{\sigma}_{\pm} = \dfrac{1}{2}(\hat{\sigma}_x \pm i\hat{\sigma}_y)$ \cite{cottet}, we obtain, 
\begin{align}
\dfrac{H_\text{tot}}{\hbar} = \dfrac{\delta\omega}{2}\hat{\sigma}_z +  \alpha \sqrt{2\kappa_\text{em}}\hat{\sigma}_y
\label{eq:simplified_slh}
\end{align}
 Using the fact that $\Gamma_\mathrm{1D}^\mathrm{f} = 2\kappa_\text{em}$ for $\varphi_\mathrm{c} = \pi/2, \varphi_\mathrm{WG} = \pi/2$ (see \cref{eq:forward_gamma1D}), we obtain,
\begin{align}
\dfrac{H_\text{tot}}{\hbar} & = \dfrac{\delta\omega}{2}\hat{\sigma}_z +  \dfrac{\Omega_\mathrm{R}}{2}\hat{\sigma}_y   \label{eq:Rabi_Hamiltonian}
\\
\Omega_\mathrm{R} & = 2\alpha\sqrt{\Gamma_\text{1D}^\mathrm{f}} \nonumber 
\end{align}
The master equation for the chiral atom can then be written as 
\begin{align}
\dot{\rho} = -\dfrac{i}{\hbar}[H_\text{tot}, \rho] + \mathcal{L}\rho
\label{eq:master-equation}
\end{align}
where the Liouvillian $\mathcal{L}$ is given by \cite{cottet, Mirhosseini2019May}, 
\begin{align}
    \mathcal{L} = (\nth + 1) \Gamma_1 \mathcal{D}[\hat{\sigma}_{-}]\rho + \nth \Gamma_1 \mathcal{D}[\hat{\sigma}_{+}]\rho + \dfrac{\Gamma_{\phi}}{2} \mathcal{D}[\hat{\sigma}_z] \rho. 
\end{align}
where $\Gamma_1$ is the total energy relaxation rate of the qubit at zero temperature, $\Gamma_{\phi}$ is the pure dephasing rate, and $\nth = 1/(e^{\hbar\omega/k_\mathrm{B}T}-1)$ is the thermal occupation of the bath. The total energy decay rate is the sum of the radiative decay rate to the waveguide $\Gamma_\text{1D}^\mathrm{f,b}$ and energy decay to loss channels $\Gamma_\mathrm{loss}$. Here, $\Gamma_\mathrm{loss}$ includes radiative decay of other emitter sidebands to the waveguide and radiative decay of the qubit to channels other than the waveguide, such as due to coupling to two-level systems (TLS) and dielectric loss. The decoherence rate of the qubit is given by $\Gamma_2 = \Gamma_1/2 + \Gamma_\phi$. The Lindblad operator is defined in its standard form as
\begin{align}
    \mathcal{D}[X]\rho = X\rho X^\dagger - \dfrac{1}{2} X^\dagger X \rho - \dfrac{1}{2}\rho X^\dagger X .
\end{align}

\section{Limits on chirality}
\label{appendix:chirality_bound}

\subsection{Extraction of directionality ratio ($\eta_\text{d}$)}

\begin{figure}[htbp]
\centering
\includegraphics[width=1\linewidth]{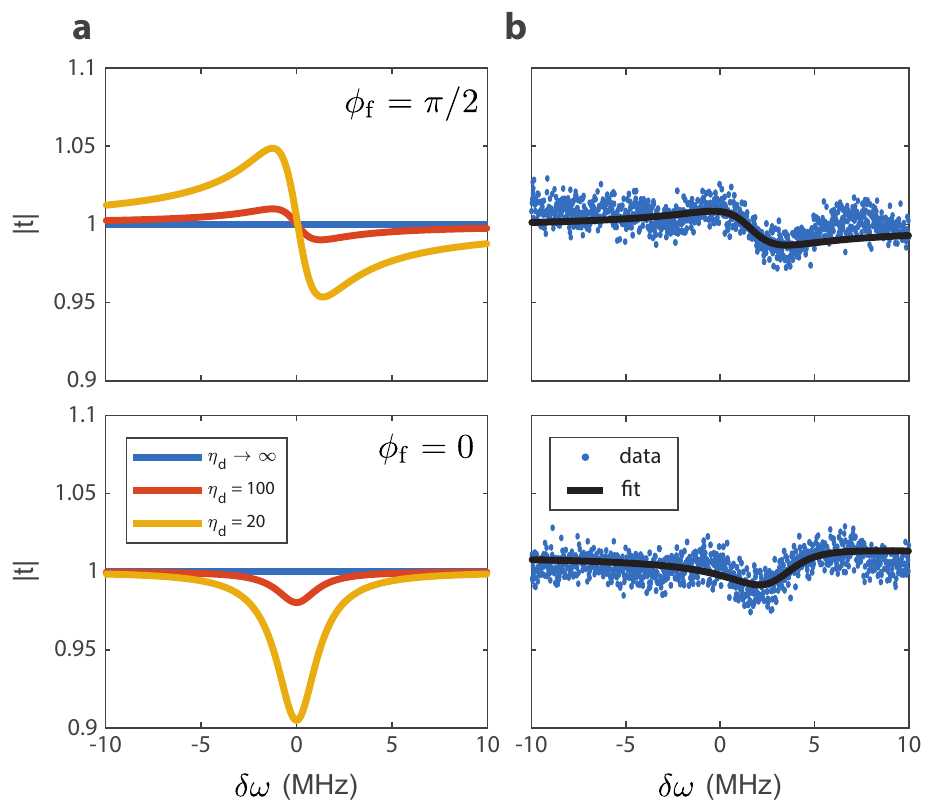}
\caption{\textbf{Extraction of directionality ratio.} (a) Theoretical transmission traces for backward excitation of the waveguide, for $\phi_\mathrm{f} = \pi/2$ and $\phi_\mathrm{f}= $0. Here, we set $\Gamma_\mathrm{1D}^\mathrm{f} =$ 2.5 MHz, $\Gamma^\prime=0$, and vary $\Gamma_\mathrm{1D}^\mathrm{b} =$ 0, 25 kHz, or 125 kHz. (b) Representative transmission traces show background fluctuations when exciting the qubit from the waveguide in the backward direction.
}


\label{fig:chirality_bound}
\end{figure}

In our device, the atom-waveguide coupling is varied by changing the emitter-coupler detunings. At each maximal chirality setting, the directionality ratio ($\eta_\text{d} = \Gf/\Gb)$ is extracted from transmission traces obtained by exciting the waveguide from the forward and backward directions. We apply non-linear least squares fits to the transmission traces according to \cref{eq:input-out-single-sided-fano}, given below. 
\begin{align}
    t = 1 - \dfrac{\Gamma_\text{1D}^\mathrm{f(b)}e^{i\phi_\mathrm{f}}}{i\delta\omega + \dfrac{\Gamma_\text{tot}}{2}}
    \label{eq:input-out-single-sided-fano}
\end{align}
\cref{eq:input-out-single-sided-fano} is obtained from \cref{eq:input-out-single-sided} by the inclusion of a `Fano' parameter, $e^{i\phi_\mathrm{f}}$, to account for asymmetric lineshapes \cite{probst2015}. The external coupling to forward (backward) waveguide modes is given by $\Gamma^\mathrm{f (b)}_\mathrm{1D}\cos(\phi_\mathrm{f})$. Detuning is given by $\delta\omega$, and total decay rate is given by $\Gamma_\mathrm{tot}$. Confidence intervals for the fitted atom-waveguide coupling rates ($\Gf, \Gb$) are used to obtain the final 95\% confidence bounds on directionality, making use of the uncorrelated non-central normal ratio distribution \cite{Hinkley1969Dec}.

Small values of backward atom-waveguide coupling make extraction of $\Gb$ challenging. For a perfect chiral atom, $\Gb = 0$. This implies that transmission $t \rightarrow 1$ (see \cref{eq:input-out-single-sided-fano} and \cref{fig:chirality_bound}a); the atom becomes invisible to photons propagating in the backward direction. Similarly, for small values of $\Gb$, the backward transmission trace approaches unity. As a result, the extraction of atom-waveguide coupling becomes susceptible to small variations in the transmission background. We observe such variations (`ripples') in backward transmission traces, which are shown in \cref{fig:chirality_bound}b. These `ripples' translate to larger uncertainties in extracting the near-zero backward emission rates at the points of maximum chirality.

We emphasize that this parasitic effect primarily compromises our characterization method for bounding the chirality ratio and not necessarily the directionality of the artificial atom. A more sensitive measurement scheme in which the emitter is driven in the forward direction while simultaneously measuring the backward scattered power will likely put tighter bounds on the backward emission (equivalent to a larger directionality ratio). These measurements were not possible in our dilution fridge at the time of performing this experiment. 

We additionally observe that `ripples' of the transmission trace backgrounds fluctuate in time, which we attribute to two-level system (TLS) defects. Such defects may couple to emitter or coupler qubits \cite{Muller2019Oct}. An example of a TLS defect is indicated in \cref{fig:tuning_curves}c. Measurements over longer periods of time have a larger chance of capturing these fluctuations, which manifests as slight dispersive changes in the transmission profile. 

\subsection{Phase stability of the coupler drives}
To determine an upper bound on experimentally achievable chirality, we characterize the phase stability of the microwave source used for coupler modulation (Rohde and Schwarz SMB 100A). The microwave source output is measured with a VNA in zero-span mode, and phase fluctuations are measured over 10 hours. The variance in phase over this duration is $\langle d\varphi^2  \rangle = 4.9$ deg$^2$. Using \cref{eq:forward_gamma1D} and \cref{eq:backward_gamma1D}, and setting $\varphi_\mathrm{WG} = \pi/2$, we have a directionality ratio $\eta_{\text{d}} = \frac{1+\text{sin}(\varphi_\mathrm{c})}{1-\text{sin}(\varphi_\mathrm{c})}$. Taylor expansion about $\varphi_\mathrm{c}$ yields $\eta_{\text{d}} \approx \frac{4}{d\varphi_\mathrm{c}^2}$. Treating the two source phases as independent random variables, we bound chirality as $\eta_{\text{d}} \approx \frac{2}{\langle d\varphi^2 \rangle}$. This yields an upper chirality bound of $\eta_{\text{d}} = 1.3 \times10^3$. Phase variability of the microwave source is potentially caused by temperature fluctuations in the measurement environment.

\section{Analysis of the decoherence sources}
\label{appendix:waveguide-temp}
\begin{figure}[t!]
\centering
\includegraphics[width=0.5\textwidth]{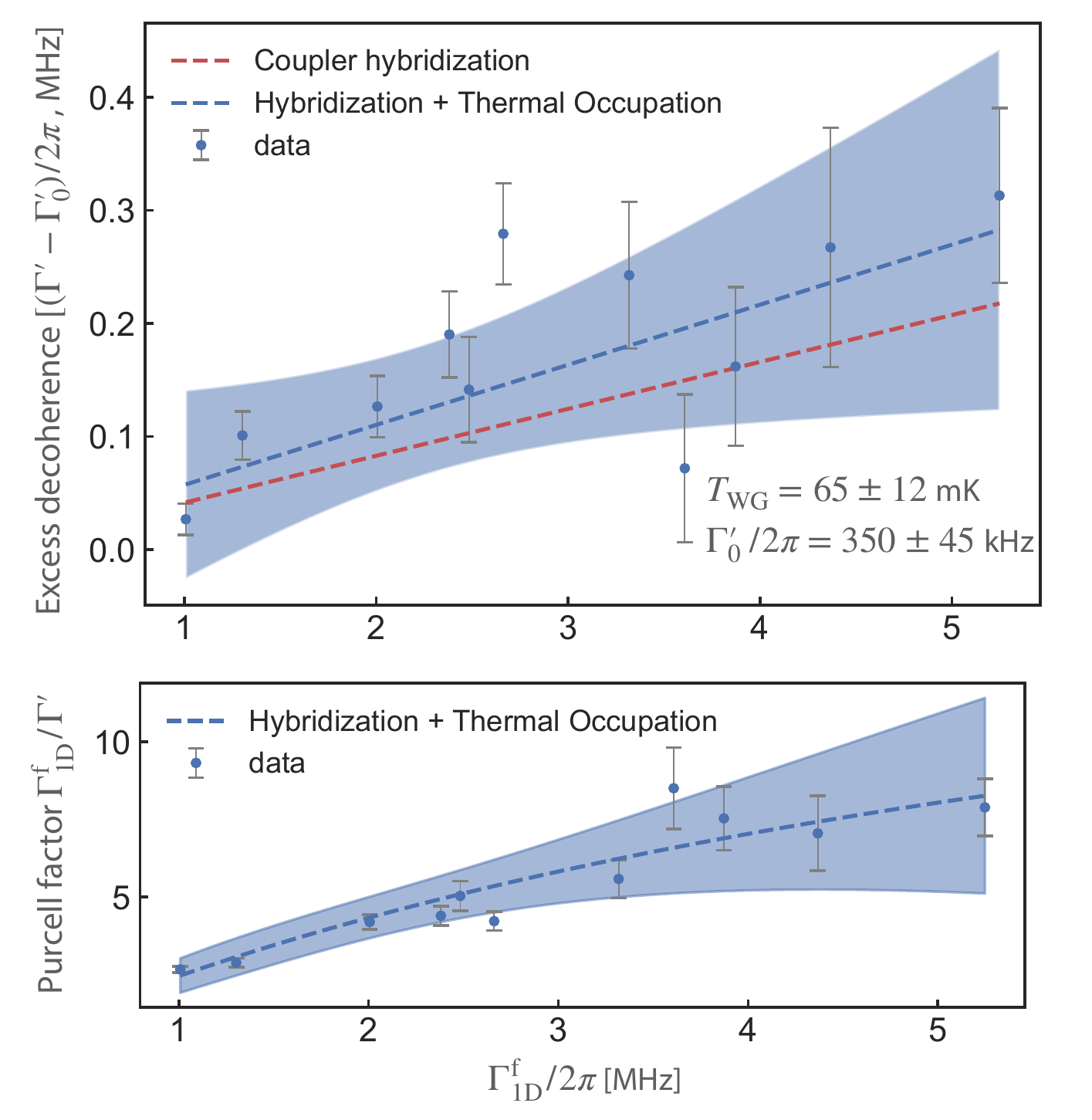}
\caption{\textbf{Emitter decoherence} (a) Measured excess decoherence ($\Gamma^\prime - \Gamma^\prime_0$) for increasing $\Gf$. The increase in decoherence can be attributed to a combination of emitter-coupler hybridization (dotted red line) and thermal occupation of the waveguide (dotted blue line). From this data, we estimate a waveguide temperature of $T_\mathrm{WG}$ = 65 $\pm$ 12 mK. The offset $\Gamma^\prime_0/2\pi$ = 350 $\pm$ 45 kHz corresponds to the decoherence rate in the limit of weak emitter-coupler hybridization and small $\Gf$. (b) Measured Purcell factor as a function of $\Gf$. The dashed blue line and shadow denote the calculated Purcell factor after accounting for emitter-coupler hybridization and finite waveguide temperature. Shadows correspond to a 95\% confidence interval from uncertainty in the waveguide temperature model.} 
\label{fig:Waveguide-Temperature}
\end{figure}



  Our measured $\beta$-factor ($\beta = \Gamma_\mathrm{1D}^f/(\Gamma_\mathrm{1D}^\mathrm{f} + \Gamma_\mathrm{1D}^\mathrm{b}+ \Gamma^{\prime}))$ is set by the intrinsic decoherence rate $\Gamma^{\prime}$ of the chiral atom. A related commonly used figure of merit for light-matter interaction is the Purcell factor, which is defined as the ratio of the emission rate of the atom to the desired waveguide modes ($\Gamma^\mathrm{f}_\text{1D}$) to the intrinsic decoherence rate $\Gamma^\prime$ of the atom \cite{paulisch2016}. The intrinsic decoherence rate is given by $\Gamma^\prime = 2\Gamma_2 - \Gamma_\text{1D} = \Gamma_\mathrm{loss} + 2\Gamma_{\phi}$. Here, $\Gamma_\mathrm{loss}$ includes non-radiative loss from various sources such as dielectric loss and coupling to two-level systems (TLS) as well as parasitic radiative decay of spurious modulation sidebands to the waveguide. For vanishingly small $\Gb$, the Purcell factor is given by $\beta/(1-\beta)$. The measured Purcell factors for our chiral atom are shown in \cref{fig:Waveguide-Temperature}. These measurements are performed at the phase settings corresponding to maximum chirality ($\varphi_c = \pi/2$), such that the atom-waveguide coupling is dominantly in the forward direction ($\Gb/\Gf \rightarrow 0$). We measure a maximum Purcell factor of 8 and find that both the $\beta$ factor and the Purcell factor saturate at large $\Gf$ due to a corresponding increase in $\Gamma^\prime$.

The observed increase in $\Gamma^{\prime}$ with increasing $\Gamma_\mathrm{1D}^\mathrm{f}$ can be partially explained by the decoherence from the couplers. The tunable couplers operate far away from their flux sweet spot and  experience significant dephasing from flux noise. As a result, increasing $\Gamma_\mathrm{1D}^\mathrm{f}$ by reducing the frequency detuning between the emitter and the couplers results in an increase in the emitter decoherence from a weak hybridization with the coupler modes. Based on the measured decoherence rate of the couplers, we estimate that $\sim$ 220 kHz of emitter decoherence can be attributed to this source at the maximum value of $\Gf$ (\cref{fig:Waveguide-Temperature}). This source of decoherence can potentially be mitigated by using tunable couplers fabricated using SQUID loops consisting of asymmetric Josephson junctions \cite{hutchings2017a}. 

In addition to emitter-coupler hybridization, the finite thermal occupation of the waveguide can also lead to excess decoherence at larger values of $\Gamma_\mathrm{1D}^\mathrm{f}$ \cite{Mirhosseini2019May}. The master equation \cref{eq:master-equation} can be solved for a mean thermal occupation $\nth$ in the waveguide to obtain the thermally-enhanced decay rate $\Gamma_\text{1}^\text{th} = (2\nth + 1)\Gamma_{1}$ and decoherence rate $\Gamma_\text{2}^\text{th} = \Gamma_1^\text{th}/2 + \Gamma_{\phi}$, where $\Gamma_1$ is the relaxation rate of the emitter at zero temperature \cite{Mirhosseini2019May}. The thermally-enhanced intrinsic decoherence rate can then be obtained as $\Gamma^{\prime} = 2\Gamma_{2}^\text{th} - \Gamma_\mathrm{1D}$. Assuming $\Gamma_2^\text{th} \approx \Gamma_1^\text{th}/2$, we can write the intrinsic decoherence rate $\Gamma^{\prime}$ as, 
\begin{align}
    \Gamma^{\prime} & = \Gamma_{1}^\text{th} - \Gamma_\text{1D} \nonumber \\
    & = (2\nth + 1)\Gamma_1 - \Gamma_\text{1D} \nonumber \\
    & = 2\nth(\Gamma_\text{1D} + \Gamma^{\prime}_0) + \Gamma^\prime_0
    \label{eq:waveguide_temp}
\end{align}
where $\Gamma^{\prime}_{0}$ is the internal dissipation rate of the emitter at zero temperature ($\nth = 0$) and $\Gamma_1 = \Gamma_\text{1D} + \Gamma^\prime_0$. From fit to the data, we obtain a waveguide temperature $T_\mathrm{WG}$ = 65 $\pm$ 12 mK and $\Gamma^{\prime}_0/2\pi$ = 350 $\pm$ 45 kHz. Better thermalization of the waveguide can be obtained using thin film microwave attenuators \cite{yeh2017}. Eliminating the dependence of $\Gamma^{\prime}$ on $\Gamma_\mathrm{1D}^\mathrm{f}$ (e.g. by using asymmetric junction SQUIDS and cryogenic attenuators) translates to a 2x improvement in the measured maximum Purcell factor in our experiment.



\begin{figure}[t!]
\centering
\includegraphics[width=0.5\textwidth]{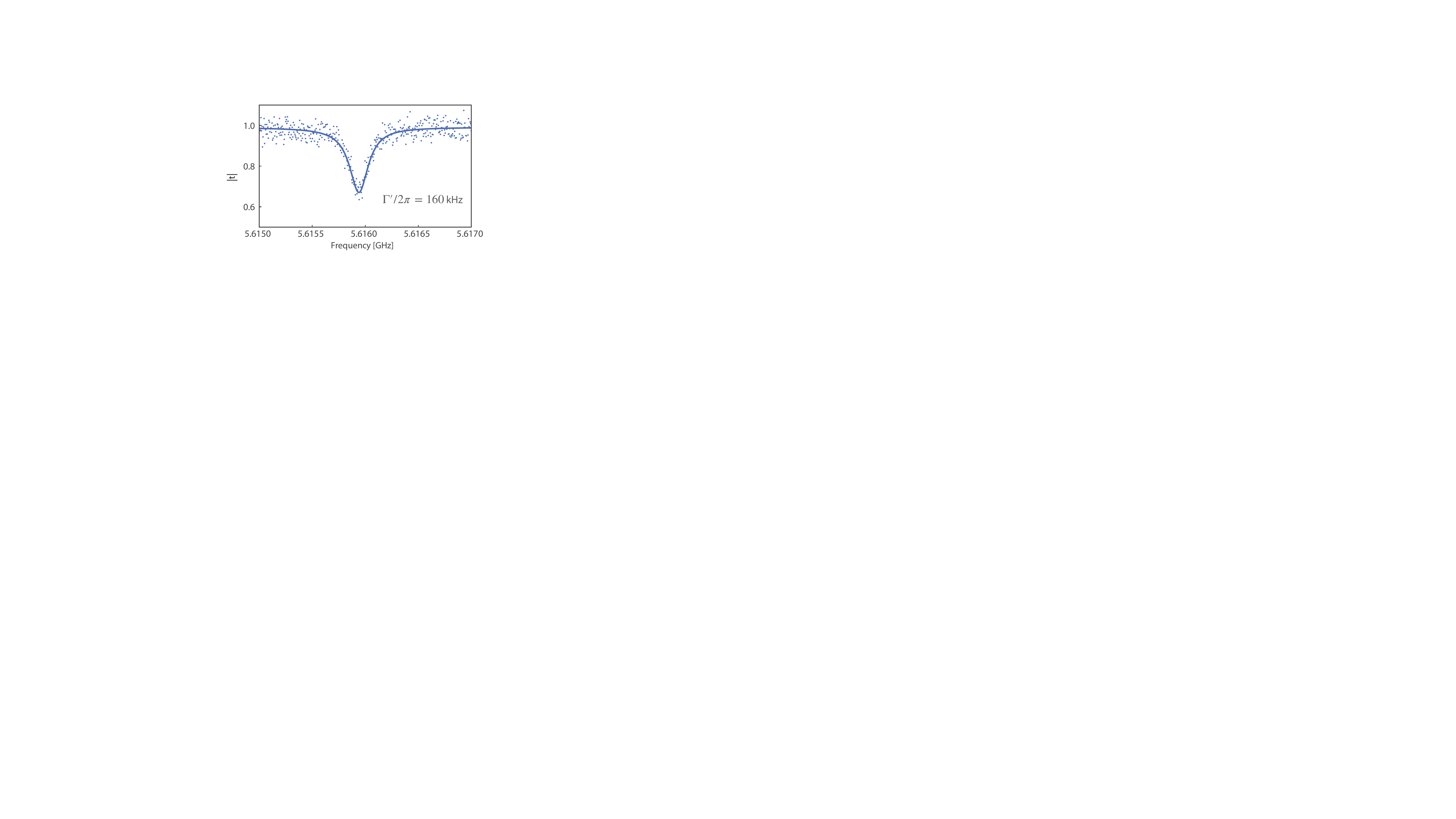}
\caption{\textbf{Characterization of the emitter baseband:} Measured transmission $|t|$ response of the emitter baseband with all RF modulation turned off. We tune the left coupler bias such that we get sufficient external coupling for the baseband to be visible when probed with a weak microwave tone via the waveguide. We measure the baseband frequency to be 5.616 GHz. The emitter baseband is undercoupled, with an internal decay rate $\Gamma^\prime$ = $2\pi \times 160$ kHz and an external coupling rate $\kappa_\mathrm{e}$ = $2\pi \times 76$ kHz.}
\label{fig:emitter_baseband}
\end{figure}

 We highlight the difference between the intrinsic decoherence rate of the chiral sideband ($\Gamma^\prime_0/2\pi$ = 350 $\pm$ 45 kHz for small values of $\Gamma_\mathrm{1D}^\mathrm{f}$), with that of the emitter baseband when all the modulation drives are off (measured to be $\Gamma^\prime$ = $2\pi \times 160$ kHz, see \cref{fig:emitter_baseband}). The difference between these two values may indicate the presence of energy leakage to spurious sidebands. Such parasitic radiative decays can be suppressed using a more aggressive sideband filtering scheme, such as replacing the filter cavities in our experiment with a dispersion-engineered metamaterial waveguide. The complete elimination of these decay channels in our experiment translates to an additional 2x improvement in the maximum measured Purcell factor. 
 
 Finally, we note that modest improvements to the intrinsic lifetime of the qubits (to 5-10 $\mu$s corresponding to an internal linewidth less than 30 kHz) translate to an additional 5x improvement to our maximum measured Purcell factors. As a result, a combination of improved coupler design, waveguide thermalization, stronger spectral filtering, and improved qubit lifetime can lead to an order of magnitude improvement in Purcell factors, resulting in $\Gf/\Gamma^\prime > 100$. Such Purcell factors have been achieved in non-chiral waveguide QED systems based on superconducting qubits \cite{ferreira2022, zanner2022}.

\label{appendix:emitter-baseband}


\section{Power broadening and resonance fluorescence}
\label{appendix:mollow}
\begin{figure}[t!]
\centering
\includegraphics[width=0.5\textwidth]{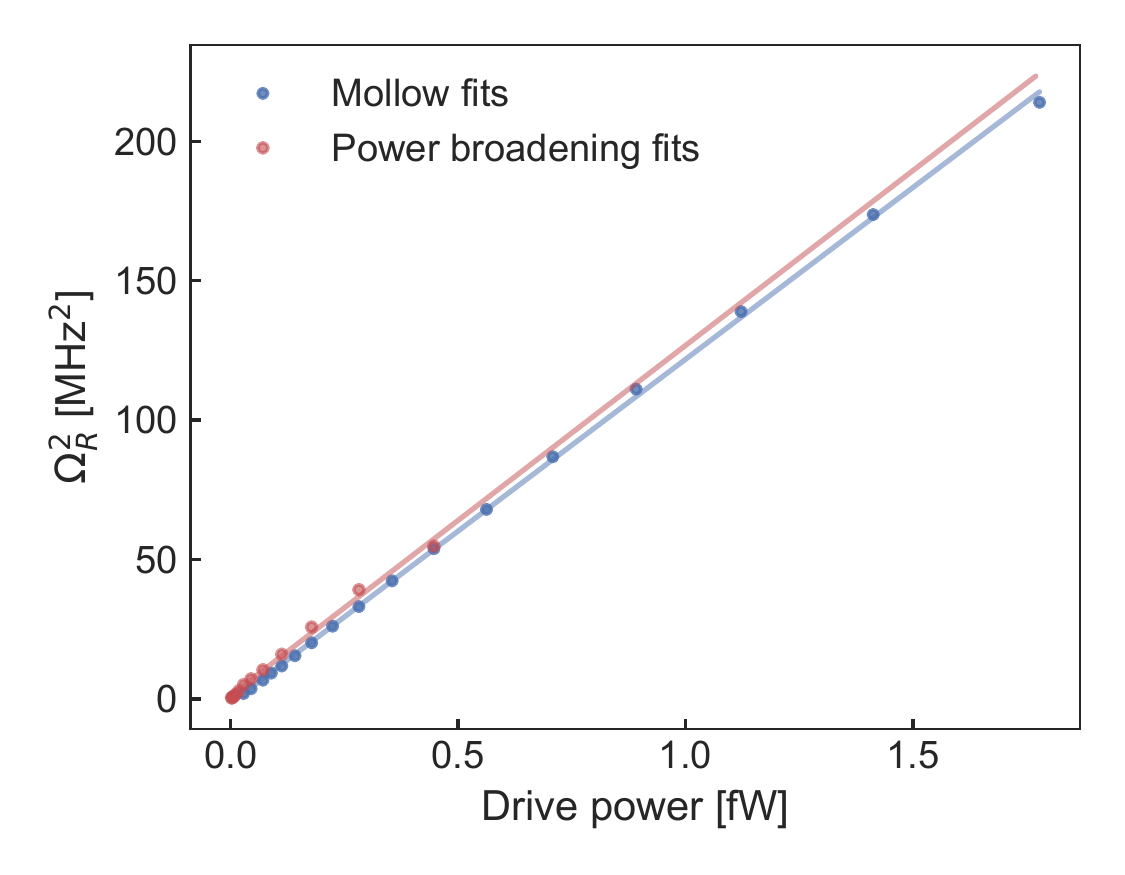}
\caption{\textbf{Rabi frequency vs drive power:} $\Omega_\mathrm{R}^2$ vs drive power obtained from least-squares fitting to the Mollow triplet data (blue) and power broadening data (red). Solid lines are linear fits to the model $\Omega_\mathrm{R}^2 = 4\Gamma_\mathrm{1D}P_\mathrm{in}/\hbar\omega_\mathrm{ge}$. From these fits, we obtain a slope of $m = 123 \pm 2\text{ MHz}^2/\mathrm{fW}$ for the Mollow data and a slope of $m = 125 \pm 6\text{ MHz}^2/\mathrm{fW}$ for the power broadening data. Uncertainties on the slope represent 95\% CI obtained from the least-squares fitting routine.}
\label{fig:Rabi-Mollow-vs-PowerBroadening}
\end{figure}

{
\renewcommand{\arraystretch}{1}
\begin{table}
\centering
\resizebox{1\linewidth}{!}{
\begin{tabular}{lcc}
\toprule
& $\Gamma_1/2\pi$ [MHz] & $\Gamma_2/2\pi$ [MHz] \\
\midrule
Mollow fits & 1.34 $\pm$ 0.15 & 0.72 $\pm$ 0.08 \\
Power broadening fits & 1.35 $\pm$ 0.03 & 0.67 $\pm$ 0.1 \\
\bottomrule
\end{tabular}
}             
\caption{Summary of fit parameters from the resonance fluorescence and coherent power broadening measurements shown in \cref{fig:fig4} of the main text. For these measurements, we independently measure the emitter-waveguide coupling rate to be $\Gamma_\text{1D}^\mathrm{f}/2\pi$ = 1 MHz.}
\label{tab:fit_params}
\end{table}
}
Here, we analyze the behavior of the chiral atom under strong drives. At zero-temperature ($\nth$ = 0),  \cref{eq:master-equation} can be solved to give the steady-state solutions of the Bloch equations \cite{cottet}. Using the relations $\langle \sigma_x \rangle = 2\text{Re}(\rho_\text{eg}), \langle \sigma_y \rangle = -2 \text{Im}(\rho_\text{ge})$ and $\langle \sigma_z \rangle = \rho_\text{ee} - \rho_\text{gg}$. We obtain, 
\begin{align}
\langle {\dot\sigma_x} \rangle &= \Omega_\mathrm{R} \langle \sigma_z \rangle - \delta\omega \exc{\sigma_y} - \Gamma_2 \exc{\sigma_x}\\
\langle {\dot\sigma_y} \rangle &= \delta\omega \exc{\sigma_x} - \Gamma_2\exc{\sigma_y}\\
\langle {\dot\sigma_z} \rangle &= -\Omega_\mathrm{R} \exc{\sigma_x} - \Gamma_1 (1 + \exc{\sigma_z}) 
\label{eq:evolution_equations}
\end{align}
where we have used the commutation relations $[\sigma_x, \sigma_y] = 2i\sigma_z$, $[\sigma_y, \sigma_z] = 2i\sigma_x$ and $[\sigma_z, \sigma_x] = 2i\sigma_y$ to evaluate \cref{eq:master-equation}. In steady state, we obtain the solutions \cite{cottet}, 
\begin{align}
     \exc{\sigma_x} & = \dfrac{-\Gamma_1 \Gamma_2 \Omega_\mathrm{R}}{\Gamma_1(\Gamma_2^2 + \delta\omega^2) + \Gamma_2 \Omega_\mathrm{R}^2} \\
    \exc{\sigma_y} & = \dfrac{-\Gamma_1 \delta\omega\Omega_\mathrm{R}}{\Gamma_1(\Gamma_2^2 + \delta\omega^2) + \Gamma_2 \Omega_\mathrm{R}^2} \\
    \exc{\sigma_z} & = -1 + \dfrac{\Gamma_2\Omega_\mathrm{R}^2}{\Gamma_1(\Gamma_2^2 + \delta\omega^2) + \Gamma_2 \Omega_\mathrm{R}^2}.
\end{align}
Using the input-output relations shown in  \cref{eq:SLH-input-output} and the relation $\hat{\sigma}_{-} = \frac{1}{2}(\hat{\sigma}_x - i\hat{\sigma}_y)$, we obtain the coherent response of the driven chiral qubit, 
\begin{align}
t = 1 - \dfrac{\Gamma_\text{1D}^\mathrm{f}\Gamma_1(\Gamma_2 - i\delta\omega)}{\Gamma_1(\Gamma_2^2 + \delta\omega^2) + \Gamma_2\Omega_\mathrm{R}^2}
\label{eq:power-broadening-proof}
\end{align}
where we have used $\Gamma_\text{1D}^\mathrm{f} = 2\kappa_\text{em}$ for $kd = \pi/2 \text{ and } \varphi_\mathrm{c} = \pi/2$. 
The power spectrum of the output radiation field is given by \cite{koshino2012}, 
\begin{align}
    S(\omega) = \text{Re} \int_{0}^{\infty}  \dfrac{d\tau}{\pi} e^{i\omega t} \langle a_\text{out}^\dagger(t) a_\text{out}(t + \tau)\rangle
    \label{eq:weiner-khinchin}
\end{align}
For the case of perfect chirality ($\varphi_\mathrm{c} = \pi/2$, $kd = \pi/2$), the output field is given by (see \cref{eq:SLH-input-output}, 
\begin{align}
    \hat{a}_\text{out}(t) = \hat{a}_\text{in}(t) + \sqrt{\Gamma_\mathrm{1D}^\mathrm{f}}\sigma_{-}(t).
\end{align}
\cref{eq:weiner-khinchin} contains both the coherent and incoherent part of the emission spectrum. The incoherent part can be evaluated to be \cite{koshino2012},

\begin{align}
\begin{split}\label{eq:mollow_appendix}
S(\omega) = \dfrac{1}{2\pi} & \dfrac{\hbar \omega_0 \Gamma_\text{1D}^\mathrm{f}}{4} \left (\dfrac{\Gamma_\mathrm{s}}{\left(\delta\omega + \Omega_\mathrm{R}\right)^2 + \Gamma_\mathrm{s}^2}  \right. \\ & + 
\left. \dfrac{2\Gamma_2}{\delta\omega  + \Gamma_2^2} + 
\dfrac{\Gamma_\mathrm{s}}{\left(\delta\omega - \Omega_\mathrm{R}\right)^2 + \Gamma_\mathrm{s}^2} \right)
\end{split}
\end{align}
where $2\Gamma_\mathrm{s} = \Gamma_1 + \Gamma_2$ is the full-width at half-maximum of the Mollow sidebands.

We choose a setting with $\Gamma_\text{1D}^\mathrm{f}/2\pi$ = 1 MHz for the resonance fluorescence and coherent power broadening measurements. The emitter-coupler detuning is chosen to be $\delta_\mathrm{CE}/2\pi$ = 105 MHz to avoid exciting the couplers under strong drives. We perform non-linear least-squares fitting of the measured power-broadened coherent response of the chiral emitter to the model in \cref{eq:power-broadening-proof} and the resonance fluorescence data to the model in \cref{eq:mollow_appendix}. $\Gf/2\pi = 1 \pm 0.1$ MHz was independently obtained using VNA measurements at low powers and fixed while performing the fits. The Rabi frequencies $\Omega_\mathrm{R}$ obtained from these fits are shown in \cref{fig:Rabi-Mollow-vs-PowerBroadening}. For a chiral atom, we expect $\Omega_\mathrm{R} = \sqrt{4\Gf P_\mathrm{in}/\hbar\omega_\mathrm{ge}}$ (see \cref{eq:Rabi_Hamiltonian}). The drive power at the chip $P_\mathrm{in}$ in \cref{fig:Rabi-Mollow-vs-PowerBroadening} is calculated using input line attenuation calibrated using thermometry measurements \cite{joshi2022}. From linear fits to the $\Omega_\mathrm{R}^2$ vs drive power data, we obtain a slope $m = 123 \pm 2 \text{ MHz}^2/\text{fW}$ for the Mollow data and  $m = 125 \pm 6 \text{ MHz}^2/\text{fW}$ for the power broadening data, which are in agreement with each other. From measured values of $\Gf/2\pi = 1 \pm 0.1$ MHz, we expect a slope $m = \frac{4\Gf}{{(2\pi)^2\hbar\omega_\mathrm{ge}}} = 150 \pm 15 \text{ MHz}^2/\text{fW}$, which is close to the values obtained from \cref{fig:Rabi-Mollow-vs-PowerBroadening}. We attribute the discrepancy to uncertainty in the line attenuation calibration used to obtain the drive power ($\sim$ 15\%) \cite{joshi2022}. 

The energy decay rate $\Gamma_1$ and the decoherence rate $\Gamma_2$ obtained from these fits are summarized in \cref{tab:fit_params}, showing good agreement with each other. From the values of $\Gamma_1$ and $\Gamma_2$ obtained from the resonance fluorescence data, we obtain a small pure dephasing rate $\Gamma_{\phi}/2\pi \approx 50$ kHz for this setting. We also obtain the internal dissipation rate for the emitter qubit $\Gamma^\prime = \Gamma_1 - \Gf = 2\pi \times 364$ kHz at this setting. 



\section{Bounding chirality of the $\ef$ transition}
\label{appendix:EF}


The chirality of the $e-f$ transition is difficult to extract because variations with respect to $\varphi_c$ in $\Gf$ and $\Gb$ for $e-f$ are confounded with an analogous variation in atom-waveguide coupling for the $g-e$ transition. Hence, we bound the $e-f$ chirality by examining the atom-waveguide coupling through each dissipation port. 



We extract atom-waveguide couplings from fits to transmission traces obtained from two-tone spectroscopy. Using a master equation treatment \cite{koshino2012} of a three-level system, we obtain the input-output relations and the corresponding complex transmission coefficient $t$ for the $e-f$ transition under chiral and bidirectional settings, with the $g-e$ transition under a strong continuous drive. For the sake of brevity, full analytical expressions are omitted. Fits to analytical expressions are supplemented with corresponding master equation simulations performed using QuTiP \cite{johansson2013}.


In our experiment, the $e-f$ transition exhibits imbalanced atom-waveguide coupling between the two dissipation ports ($\kappa_\mathrm{em}^\mathrm{l} \neq \kappa_\mathrm{em}^\mathrm{r}$).  
For this case, expressions for $\Gf$ and $\Gb$ can be obtained following \cref{appendix:input-output}.
\begin{align}
    \Gamma_\mathrm{1D,e-f}^\mathrm{f,b} = \frac{\kappa_\mathrm{em}^\mathrm{l}  +\kappa_\mathrm{em}^\mathrm{r}}{2} \pm \sqrt{\kappa_\mathrm{em}^\mathrm{l} \kappa_\mathrm{em}^\mathrm{r}}
    \label{eq:imbalance_gamma}
\end{align}

Transmission traces are fit in the case of a continuous drive tone at frequency $\omega_\mathrm{ge}$. Values obtained from fits are used in \cref{eq:imbalance_gamma} to obtain $\Gamma_\mathrm{1D,ef}^\mathrm{f}/2\pi$ = 2.4 MHz, and $\Gamma_\mathrm{1D,ef}^\mathrm{b}/2\pi$ = 0.2 MHz, corresponding to a directionality ratio of $\eta_\mathrm{d}$ = 12. In addition, these fits yield $\Gamma_\mathrm{ef}^\prime/2\pi$ = 0.75 kHz, corresponding to $\Gamma_\mathrm{tot, ef}/2\pi$ = 3.15 MHz. An additional fit to the $g-e$ transition measured independently on the VNA yields $\Gamma^\prime_\mathrm{ge}/2\pi$ = 0.7 MHz and $\Gamma_\mathrm{tot, ge}/2\pi$ = 1.3 MHz. 

In principle, with pulsed excitation of the $g-e$ transition, the parameters obtained above are sufficient to observe a chiral response with strong coupling for the e-f transition ($\Gamma_\mathrm{1D, ef}^\mathrm{f} > 0.5(\Gamma_\mathrm{tot, ef} + \Gamma_\mathrm{tot, ge})$). However, as the $g-e$ transition is not protected from waveguide decay, we use a readout pulse of 120 ns duration to rapidly probe the $e-f$ transition in the pulsed spectroscopy measurement. To operate in the quasi-cw regime, it is necessary to use a dispersion-engineered metamaterial waveguide, such that the $e-f$ transition falls in the passband while the $g-e$ transition falls outside and remains protected from radiative decay to the waveguide \cite{ferreira2022}. Such a device architecture can be used to realize conditional phase gates on itinerant photons \cite{Kono2018Jun,Besse2018Apr, Guimond2020Mar}. 

\section{Rabi oscillations}
\label{appendix:Rabi-oscillations}
As our device is not equipped with a readout resonator, we observe Rabi oscillations by directly driving the qubit via the waveguide with a Gaussian pulse of variable duration $\tau_\text{P}$ and measuring the in-quadrature component of the qubit Bloch vector. We assume a coherent drive along $\sigma_y$, and starting with a qubit ground state $\exc{\sigma_z} (t=0) = -1$. For a resonant drive at $\delta\omega = 0$, with $\Omega_\mathrm{R} \ge |\Gamma_1 - \Gamma_2|/2$, the qubit state is given by 

\begin{align}
\begin{split}
\exc{\sigma_x}(t = \tp) &= x_{\infty} - \bigg(\frac{\Gamma_\mathrm{R} x_\infty -\Omega_\mathrm{R}}{\nu_\mathrm{R}}\sin{(\nu_\mathrm{R} \tp)} \\
&  \quad + x_\infty \cos{(\nu_\mathrm{R} \tp)}\bigg) \exp({-\Gamma_\mathrm{R} \tp})
\label{eq:sx}
\end{split}
\end{align}

\begin{align}
\begin{split}
\exc{\sigma_z}(t = \tp) &= z_{\infty} - (1+z_\infty) \bigg(\cos(\nu_\mathrm{R}\tp) \\
& \quad +\frac{\Gamma_\mathrm{R}}{\nu_\mathrm{R}}\sin(\nu_\mathrm{R} \tp)\bigg) \exp({-\Gamma_\mathrm{R} \tp})
\label{eq:sz}
\end{split}
\end{align}

The resonant drive $\delta\omega = 0$ results in $\exc{\sigma_y}(t = \tp) = 0$. Here, $\Gamma_\mathrm{R} = (\Gamma_1 + \Gamma_2)/2$ is the Rabi decay rate and $\nu_\mathrm{R} = \sqrt{\Omega_\mathrm{R}^2 - (\Gamma_1 - \Gamma_2)^2/{4}}$ is the effective Rabi oscillation frequency. The steady-state values of $\exc{\sigma_x}(t=\infty)$ and $\exc{\sigma_z}(t=\infty)$ are $x_\infty = {\Gamma_1 \Omega_\mathrm{R}}/{(\Gamma_1 \Gamma_2 + \Omega_\mathrm{R}^2)}$ and $z_\infty = -\Gamma_1 \Gamma_2/{(\Gamma_1 \Gamma_2 + \Omega_\mathrm{R}^2)}$, respectively. Rabi oscillation measurements are fitted to \cref{eq:sx} assuming no dephasing ($\Gamma_2 = \Gamma_1/2$). 

In the limit of large drives ($\Omega_\mathrm{R} \gg \Gamma_1, \Gamma_2$), $\nu_\mathrm{R} \approx \Omega_\mathrm{R}$ and $x_\infty \approx z_\infty \approx 0$. \cref{eq:sx} and \cref{eq:sz} may be simplified to

\begin{align}
\exc{\sigma_x}(t = \tp) &= \sin{(\Omega_\mathrm{R} \tp)} \exp{\left(-\Gamma_\mathrm{R}\tp\right)} 
\label{eq:rabi_tau_p}
\end{align}
\begin{align}
\exc{\sigma_z}(t = \tp) &= -\cos{(\Omega_\mathrm{R} \tp)} \exp{\left(-\Gamma_\mathrm{R}\tp\right)}.
\end{align}
 After the drive is turned off, the qubit evolves freely and decays due to energy dissipation and decoherence. The evolution equations can be obtained by substituting $\Omega_\mathrm{R} = 0$ in \cref{eq:evolution_equations}, 
\begin{align}
\exc{{\dot\sigma_x}} & = -\Gamma_2 \exc{\sigma_x} \\
\exc{{\dot\sigma_z}} & = -\Gamma_1 (1 + \exc{\sigma_z}).
\label{eq:rabi_evolution}
\end{align}
The solutions to the above equations are given by, 
\begin{align}
    \exc{\sigma_x}(t^\prime) = \exc{\sigma_x}(\tp) \exp \left({-\Gamma_2 t^\prime}\right)\\
    \exc{\sigma_z}(t^\prime) = [1 + \exc{\sigma_z}(\tp)]\exp{\left(-\Gamma_1 t^\prime\right)} - 1
\end{align}
where $\exc{\sigma_x}(\tp)$ and $\exc{\sigma_z}(\tp)$ are obtained from \cref{eq:rabi_tau_p}. As we perform readout directly via the waveguide, the signal-to-noise ratio (SNR) is limited by HEMT noise. To obtain better SNR, we, therefore, perform phase-sensitive averaging (w.r.t the Rabi drive) of the qubit emission after the drive is turned off. The component of qubit emission in-quadrature with the drive gives us $\exc{\sigma_x}$. We integrate the ring-down, resulting in a signal $\int dt^\prime \exc{\sigma_x}(\tp) \exp{\left(-\Gamma_2 t^\prime \right)}$, which is proportional to $\exc{\sigma_x}(\tp)$. 

To perform these measurements, we first generate Gaussian pulses at the intermediate frequency of 60 MHz using the Quantum Machines OPX+ module. The pulses are next up-converted to radio frequencies by combining them with a local oscillator (LO) using a mixer.  After driving the qubit with the resonant Gaussian pulse, qubit emission is down-converted with another mixer using the same LO that is used for generating the drive. The output is then demodulated, and the in-phase ($I$) and quadrature ($Q$) components of the output signal are averaged separately. Combining the two (as $I+i Q$) yields the projection of the qubit state onto the Bloch sphere $XY$ plane. The component in-quadrature with the drive maps to $\exc{\sigma_x}(\tp)$, which is shown in the measurements in the main text.

\section{Parametric qubit-waveguide coupling }
\label{appendix:CMT_Full}
\begin{figure*}[htbp]
\centering
\includegraphics[width=0.80\linewidth]{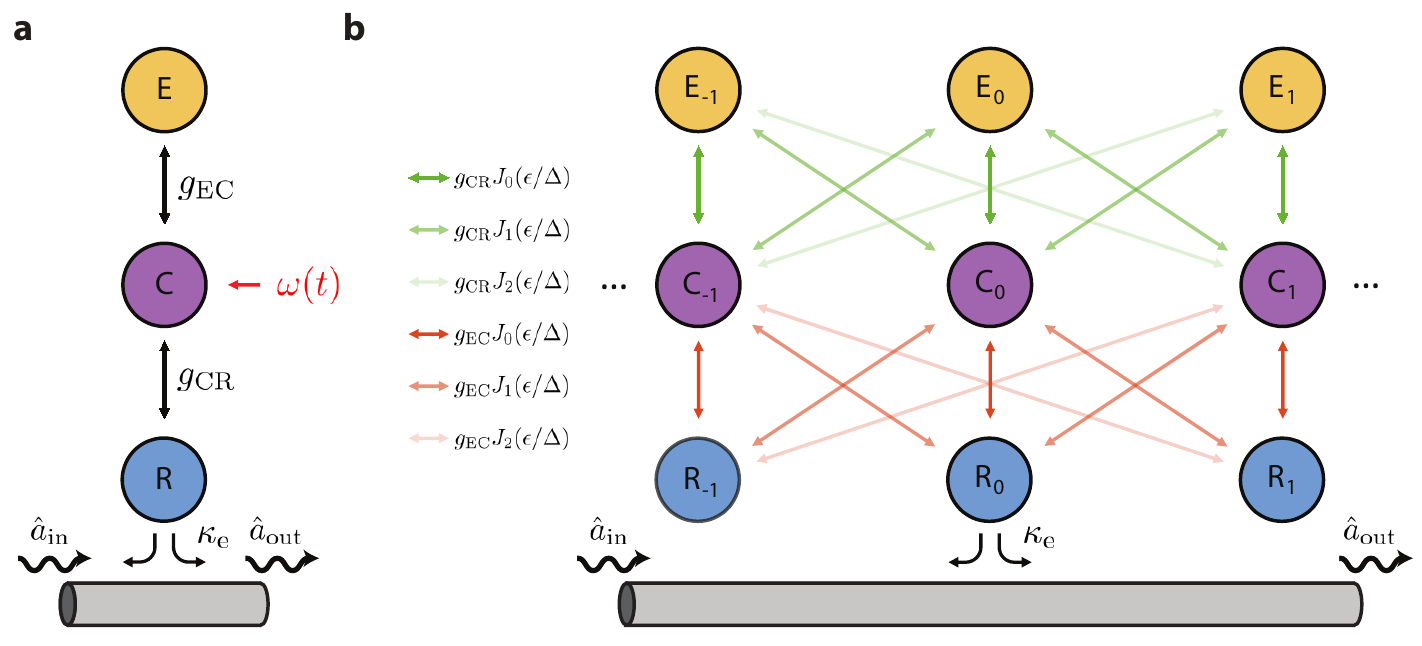}
\caption{\textbf{Parametric Waveguide Coupling Input-Output Model} (a) Schematic of tunable coupler containing three modes - an emitter, coupler, and filter cavity. The emitter-coupler and coupler-filter cavity interaction strengths are $g_\mathrm{EC}$ and $g_\mathrm{CR}$, respectively. The coupler frequency is modulated in time, and the filter cavity decays to a 1D waveguide. (b) Frequency domain picture of the tunable coupler. Frequency modulation of the coupler generates sidebands for all three modes. Effective coupling strengths between sidebands of different modes are indicated by arrows, and depend on the relative sideband order ($D$), modulation frequency ($\Delta$), and modulation amplitude ($\epsilon$). The filter cavity baseband ($R_0$) provides a decay channel to the waveguide. In experiments, the emitter's first sideband ($E_1$) is used as the chiral qubit.}
\label{fig:CMT_theory}
\end{figure*}

In our device, we generate time-harmonic coupling between the emitter qubit and a filter cavity by modulating the frequency of an intermediary mode, the coupler. This approach can be used generally to obtain non-reciprocal interactions between modes \cite{10.21468/SciPostPhysLectNotes.44}. 
Previous works have operated similar tunable couplers in the dispersive regime, when the coupler is far detuned from both the emitter and filter cavity ($\delta_\mathrm{C,E} \gg g_\mathrm{EC},  \delta_\mathrm{C,R} \gg g_\mathrm{CR}$, where $\delta_\mathrm{C,E} = \omega_\mathrm{C} - \omega_\mathrm{E}, \delta_\mathrm{C,R} = \omega_\mathrm{C} - \omega_\mathrm{R}$) \cite{Yan2018Nov, Kannan2022Mar}. Experimentally, we improve emitter-waveguide coupling by breaking the dispersive assumption. We instead allow the couplers and filter cavities to hybridize. To describe the tunable coupler in this regime, we present an input-output model for harmonic coupling between two modes by frequency modulation of an intermediary mode. The model accounts for the appearance of sidebands of the emitter as well as its effective coupling strength to sidebands of other modes. We find that the coupling strength between sidebands of distinct modes depends on the relative sideband order, the modulation frequency, and the modulation amplitude. These insights provide the means for experimental optimization of directionality and external coupling in the chiral qubit.

\subsection{Derivation of coupled mode equations}
We focus our attention on one dissipation port of the emitter. A coupler mode is directly coupled to the emitter with strength $g_\mathrm{EC}$ and directly coupled to a filter cavity mode with strength $g_\mathrm{CR}$. The filter cavity decays to photonic modes of a 1D waveguide. The coupler mode is frequency modulated. The system is shown schematically in \cref{fig:CMT_theory}a.

The full Hamiltonian in the Heisenberg picture is given as follows. We set $\hslash=1$ and assume linear cavities for simplicity.
\begin{equation}
\hat{H} = \hat{H}_\mathrm{sys} + \hat{H}_\mathrm{bath} + \hat{H}_\mathrm{int} \\
\end{equation}
\begin{equation}
\begin{split}
\hat{H}_\mathrm{sys} &= [\omega_\mathrm{C} + \epsilon \sin(\Delta t)] \hat{a}_\mathrm{C}^\dag \hat{a}_\mathrm{C} + \omega_\mathrm{R} \hat{a}_\mathrm{R}^\dag \hat{a}_\mathrm{R} + \omega_\mathrm{E} \hat{a}_\mathrm{E}^\dag \hat{a}_\mathrm{E} \\
&\quad + g_\mathrm{EC}(\hat{a}_\mathrm{E}^\dag \hat{a}_\mathrm{C} + \hat{a}_\mathrm{E} \hat{a}_\mathrm{C}^\dag) + g_\mathrm{CR}(\hat{a}_\mathrm{C}^\dag \hat{a}_\mathrm{R} + \hat{a}_\mathrm{C} \hat{a}_\mathrm{R}^\dag)
\end{split}
\end{equation}
\begin{equation}
\hat{H}_\mathrm{bath} =  \sum_{q} \omega_{q} \hat{b}_q^\dag \hat{b}_q
\end{equation}
\begin{equation}
\hat{H}_\mathrm{int} = -i\sum_{q} (f_{q} \hat{b}_{q} \hat{a}_\mathrm{R}^\dag - f_{q}^* \hat{b}_{q}^\dag \hat{a}_\mathrm{R}) 
\end{equation}
Here, $\hat{H}_\mathrm{sys}$ contains the three individual modes, the emitter-coupler interaction term, and the coupler-filter cavity interaction term. The coupler frequency is modulated at frequency $\Delta$ with amplitude $\epsilon$. $\hat{H}_\mathrm{bath}$ describes the waveguide modes, and $\hat{H}_\mathrm{int}$ describes the coupling between the filter cavity and waveguide modes. Following \cite{Clerk2010Apr}, Langevin equations are given below. 
\begin{equation}
\dot{\hat{a}}_\mathrm{R} = i\omega_\mathrm{R} \hat{a}_\mathrm{R} + \frac{\kappa_\mathrm{t}}{2}\hat{a}_\mathrm{R} + ig_\mathrm{CR} \hat{a}_\mathrm{C} - \sqrt{\frac{\kappa_\mathrm{e}}{2}} \hat{a}_\mathrm{in}
\end{equation}
\begin{equation}
\dot{\hat{a}}_\mathrm{C} = i[\omega_\mathrm{C} + \epsilon \sin(\Delta t)] \hat{a}_\mathrm{C} + \frac{\gamma_\mathrm{C}}{2}\hat{a}_\mathrm{C} + ig_\mathrm{CR} \hat{a}_\mathrm{R} + ig_\mathrm{EC} \hat{a}_\mathrm{E}
\end{equation}
\begin{equation}
\dot{\hat{a}}_\mathrm{E} = i\omega_\mathrm{E} \hat{a}_\mathrm{E} + \frac{\gamma_\mathrm{E}}{2}\hat{a}_\mathrm{E} + ig_\mathrm{EC} \hat{a}_\mathrm{C}
\end{equation}

Here, $\kappa_\mathrm{t}$ is the total decay rate of the filter cavity and $\kappa_\mathrm{e}$ is the external coupling of the filter cavity to the waveguide. The emitter and coupler do not couple directly to the waveguide, and their total decay rates $\gamma_\mathrm{E}$ and $\gamma_\mathrm{C}$ are comprised of only internal loss. We note that, in the absence of coupler frequency modulation ($\epsilon = 0$), these expressions yield electromagnetically induced transparency (EIT) phenomena.

The time dependence of the coupler frequency can be simplified by performing the following substitution.
\begin{equation}
\hat{\tilde{a}}_\mathrm{C} = \hat{a}_\mathrm{C} \mathrm{exp}[{i\frac{\epsilon}{\Delta}\cos(\Delta t)}]
\end{equation}

This is equivalent to performing a unitary transformation of the form $\hat{\tilde{a}}_\mathrm{C} = U^\dag \hat{a}_\mathrm{C} U$, where
\begin{equation}
U = \exp[-i \frac{\epsilon}{\Delta} \cos(\Delta t)\hat{a}_\mathrm{C}^\dag \hat{a}_\mathrm{C}].
\end{equation}
In the classical formulation of this problem, this transformation corresponds to an appropriate gauge  transformation \cite{Minkov2017Jul}.

Next, we make use of the Jacobi-Anger expansion, given below, to break the time dependence of the modified coupler operator into discrete harmonics. $J_n$ represents the $n$-th Bessel function of the first kind.

\begin{equation}
    \mathrm{exp}[{i \frac{\epsilon}{\Delta} \cos(\Delta t)]} = \sum_{n = -\infty}^{\infty} i^n J_n(\frac{\epsilon}{\Delta})e^{in\Delta t} 
\end{equation}

Performing the substitution then yields the following modified Langevin equations.
\begin{equation}
\begin{split}
\dot{\hat{a}}_\mathrm{R} &= i\omega_\mathrm{R} \hat{a}_\mathrm{R} + \frac{\kappa_\mathrm{t}}{2}\hat{a}_\mathrm{R} + ig_\mathrm{CR} \sum_{n} (-i)^n J_n(\frac{\epsilon}{\Delta})e^{in\Delta t} \hat{\tilde{a}}_\mathrm{C} \\
& \quad - \sqrt{\frac{\kappa_\mathrm{e}}{2}} \hat{a}_\mathrm{in}
\end{split}
\end{equation}
\begin{equation}
\begin{split}
\dot{\hat{\tilde{a}}}_\mathrm{C} &= i\omega_\mathrm{C} \hat{\tilde{a}}_\mathrm{C} + \frac{\gamma_\mathrm{C}}{2}\hat{\tilde{a}}_\mathrm{C}  
+ ig_\mathrm{CR} \sum_{l} i^l J_l(\frac{\epsilon}{\Delta})e^{il\Delta t} \hat{a}_\mathrm{R} \\
& \quad + ig_\mathrm{EC} \sum_{m} i^m J_m(\frac{\epsilon}{\Delta})e^{im\Delta t} \hat{a}_\mathrm{E}
\end{split}
\end{equation}
\begin{equation}
\dot{\hat{a}}_\mathrm{E} = i\omega_\mathrm{E} \hat{a}_\mathrm{E} + \frac{\gamma_\mathrm{E}}{2}\hat{a}_\mathrm{E} + ig_\mathrm{EC} \sum_{n} (-i)^n J_n(\frac{\epsilon}{\Delta})e^{in\Delta t} \hat{\tilde{a}}_\mathrm{C}
\end{equation}

Taking the Fourier transform then yields
\begin{equation}
\begin{split}
i \omega\hat{a}_\mathrm{R}(\omega) &= i\omega_\mathrm{R} \hat{a}_\mathrm{R}(\omega) + \frac{\kappa_\mathrm{t}}{2}\hat{a}_\mathrm{R}(\omega)\\
& \quad  + ig_\mathrm{CR} \sum_{n} (-i)^n J_n(\frac{\epsilon}{\Delta}) \hat{\tilde{a}}_\mathrm{C}(\omega-n\Delta) \\
& \quad - \sqrt{\frac{\kappa_\mathrm{e}}{2}} \hat{a}_\mathrm{in}(\omega)
\end{split}
\end{equation}
\begin{equation}
\begin{split}
i \omega\hat{\tilde{a}}_\mathrm{C}(\omega) &= i\omega_\mathrm{C} \hat{\tilde{a}}_\mathrm{C}(\omega) + \frac{\gamma_\mathrm{C}}{2}\hat{\tilde{a}}_\mathrm{C}(\omega) \\ 
& \quad + ig_\mathrm{CR} \sum_{l} i^l J_l(\frac{\epsilon}{\Delta}) \hat{a}_\mathrm{R}(\omega - l\Delta) \\
& \quad + ig_\mathrm{EC} \sum_{m} i^m J_m(\frac{\epsilon}{\Delta})\hat{a}_\mathrm{E}(\omega - m\Delta)
\end{split}
\end{equation}
\begin{equation}
\begin{split}
i\omega{\hat{a}}_\mathrm{E}(\omega) &= i\omega_\mathrm{E} \hat{a}_\mathrm{E}(\omega) + \frac{\gamma_\mathrm{E}}{2}\hat{a}_\mathrm{E}(\omega) \\
& \quad + ig_\mathrm{EC} \sum_{n} (-i)^n J_n(\frac{\epsilon}{\Delta}) \hat{\tilde{a}}_\mathrm{C}(\omega-n\Delta)
\end{split}
\end{equation}
In the frequency domain, the coupled mode equations indicate that the emitter and filter cavity at frequency $\omega$ are coupled to the coupler at frequencies $\omega + n\Delta, n\in\mathbb{Z}$, with effective coupling strength determined by the Bessel functions. In this picture, each of the emitter, coupler, and filter cavity break into a spectrum of discrete harmonics, or `sidebands.' We may displace each of the three equations in frequency by $n\Delta, n\in\mathbb{Z}$, to generate equations for each sideband of the coupler, emitter, and filter cavity. This is completed in \cref{eq:CMT_FULL}, where the coupled mode equations are arranged in a matrix-vector equation.

\def\rddots#1{\cdot^{\cdot^{\cdot^{#1}}}}

\begin{equation}
i\sqrt{\frac{\kappa_\mathrm{e}}{2}}\begin{bmatrix} \vdots \\ \textbf{0} \\ \mathbf{a_\mathbf{in}}(\omega) \\ \textbf{0} \\ \vdots 
\end{bmatrix} = 
\begin{bmatrix}
\ddots & \vdots & \vdots & \vdots & \iddots \\
\dots &
\textbf{H$_{\textbf{-1}}$} & \textbf{G$_{\textbf{1}}$} & \textbf{G$_{\textbf{2}}$} & \dots \\
\dots &
\textbf{G$_{\textbf{1}}^*$} 
& \textbf{H$_{\textbf{0}}$} & \textbf{G$_{\textbf{1}}$}
& \dots \\
\dots &
\textbf{G$_{\textbf{2}}^*$} & 
\textbf{G$_{\textbf{1}}^*$} & \textbf{H$_{\textbf{1}}$}
& \dots \\
\iddots & \vdots & \vdots & \vdots & \ddots \\
\end{bmatrix}
\begin{bmatrix} \vdots \\ \mathbf{a}(\omega-\Delta) \\ \mathbf{a}(\omega) \\ \mathbf{a}(\omega+\Delta) \\ \vdots 
\end{bmatrix}
\label{eq:CMT_FULL}
\end{equation}
We will refer to the left-hand side of this equation as the waveguide input vector. The components of the waveguide input are defined as follows. 

\begin{equation}
\textbf{0} = 
\begin{bmatrix} 0 \\ 0 \\ 0 \\ \end{bmatrix}
\end{equation}

\begin{equation}
\mathbf{a_{in}}(\omega) = 
\begin{bmatrix} 0 \\ 0 \\ \hat{a}_\mathrm{in}(\omega) \\ \end{bmatrix}
\end{equation}

The only non-zero term in the waveguide input vector corresponds to the baseband of the filter cavity, which acts as the decay pathway to the waveguide. The right-hand side is composed of the Hamiltonian matrix and sideband vector. The sideband vector components contain the emitter, coupler, and filter cavity sidebands of a single order and are given as follows.

\begin{equation}
\mathbf{a} {(\omega+n\Delta)} = 
\begin{bmatrix} \hat{a}_\mathrm{E}(\omega+n\Delta) \\ \hat{a}_\mathrm{C}(\omega+n\Delta) \\ \hat{a}_\mathrm{R}(\omega+n\Delta) \\\end{bmatrix}
\end{equation}

The Hamiltonian matrix is split into $3\times3$ sub-matrices, defined below.

\renewcommand*{\arraystretch}{1.5}
\begin{align}
    \mathbf{H}_{\boldsymbol{n}} = 
    \begin{bmatrix}
        \Delta_{\mathrm{E},n}+i\frac{\gamma_\mathrm{E}}{2} & -g_\mathrm{EC}J_{0}(\frac{\epsilon}{\Delta}) & 0 \\
        -g_\mathrm{EC}J_{0}(\frac{\epsilon}{\Delta}) & \Delta_{\mathrm{C},n}+i\frac{\gamma_\mathrm{C}}{2} & -g_\mathrm{CR}J_{0}(\frac{\epsilon}{\Delta}) \\ 
        0 & -g_\mathrm{CR}J_{0}(\frac{\epsilon}{\Delta}) & \Delta_{\mathrm{R},n}+i\frac{\kappa_\mathrm{t,R}}{2} \\
    \end{bmatrix}
\end{align}

\begin{align}
    \textbf{G}_{\boldsymbol{k}} = -(i)^k 
    \begin{bmatrix}
        0 & g_\mathrm{EC}J_{k}(\frac{\epsilon}{\Delta}) & 0 \\
        g_\mathrm{EC}J_{k}(\frac{\epsilon}{\Delta}) & 0 & g_\mathrm{CR}J_{k}(\frac{\epsilon}{\Delta}) \\ 
        0 & g_\mathrm{CR}J_{k}(\frac{\epsilon}{\Delta}) & 0 \\
    \end{bmatrix}
\end{align}
\renewcommand*{\arraystretch}{1.0}

The on-diagonal sub-matrices $\bf{H}_{\boldsymbol{n}}$ account for the resonance frequencies of the emitter, coupler, and filter cavity in a single sideband order. The $\bf{H}_{\boldsymbol{n}}$ also give the emitter-coupler and coupler-filter cavity coupling strengths between sidebands of the same order (different modes). This coupling strength is scaled by the zero-th Bessel function, $J_0$. As a result, $\epsilon = 0$ results in $J_0 = 1$, and maximal coupling between sidebands of the same order. Note that here we introduce the detuning $\Delta_{i,n} = \omega - \omega_\mathrm{E} + n\Delta$ ($i = \mathrm{E,C,R}$), which is distinct from the coupler modulation frequency, $\Delta$.

Coupling between sidebands of different orders is given by the $\bf{G}_{\boldsymbol{k}}$ sub-matrices, with $k$ dictating the order of the Bessel function which scales emitter-coupler or coupler-filter cavity interaction strength. As evidenced by the Hamiltonian matrix \cref{eq:CMT_FULL}, $k$ increases for $\bf{G}_{\boldsymbol{k}}$ further from the on-diagonal $\bf{H}_{\boldsymbol{n}}$. The relative distance between sideband orders determines $k$. For example, the emitter's $n$-th sideband and the coupler's $m$-th sideband have $k = |m-n|$, meaning the coupling strength between these sidebands contains a prefactor $J_{k}(\epsilon/\Delta)$. This scaling of interaction strengths is shown schematically in \cref{fig:CMT_theory}b. When $\epsilon = 0$ (RF drive is off), there is no coupling to any sidebands (because $J_n(0) = 0, n \neq 0$).

For a given frequency drive amplitude ($\epsilon$) and frequency modulation ($\Delta$), the relative coupling strengths of sidebands are determined by the $J_n(\epsilon/\Delta)$. For low drives $\epsilon/\Delta ~< 1$, only the Bessel functions of low order have significant magnitude. Therefore, by properly truncating the coupled mode equations (by only including coupled mode equations for sidebands of a low order), we generate a finite matrix equation that allows us to solve for the transmission of the tunable coupler. Transmission is determined by the standard 2-sided cavity input-output relation. 

\begin{equation}
    \hat{a}_\mathrm{out}(\omega) = \hat{a}_\mathrm{in}(\omega) - \sqrt{\frac{\kappa_\mathrm{e}}{2}}\hat{a}_\mathrm{R}(\omega)
\end{equation}

We may solve for $\hat{a}_\mathrm{R}(\omega)$ by inverting the Hamiltonian matrix of \cref{eq:CMT_FULL}.

\subsection{Discussion}
The derived input-output model provides several physical insights into the tunable coupler. First, by modulating the coupler frequency, we generate sidebands of all three modes. Any effective coupling between sidebands of the emitter and filter cavity are mediated by sidebands of the coupler (see \cref{fig:CMT_theory}b). Any decay to the waveguide is mediated by the filter cavity baseband. In order to maximize the coupling between an emitter/filter cavity sideband and a coupler sideband, for a given modulation frequency $\Delta$, we may vary the coupler drive amplitude $\epsilon$. A larger relative distance between sideband order ($k$) will require increased drive amplitude $\epsilon$ to optimize coupling. This is because maxima of higher order Bessel functions occur at larger values of ${\epsilon}/{\Delta}$. 

\subsection{External coupling}

\begin{figure}
\centering
\includegraphics[width=0.9\linewidth]{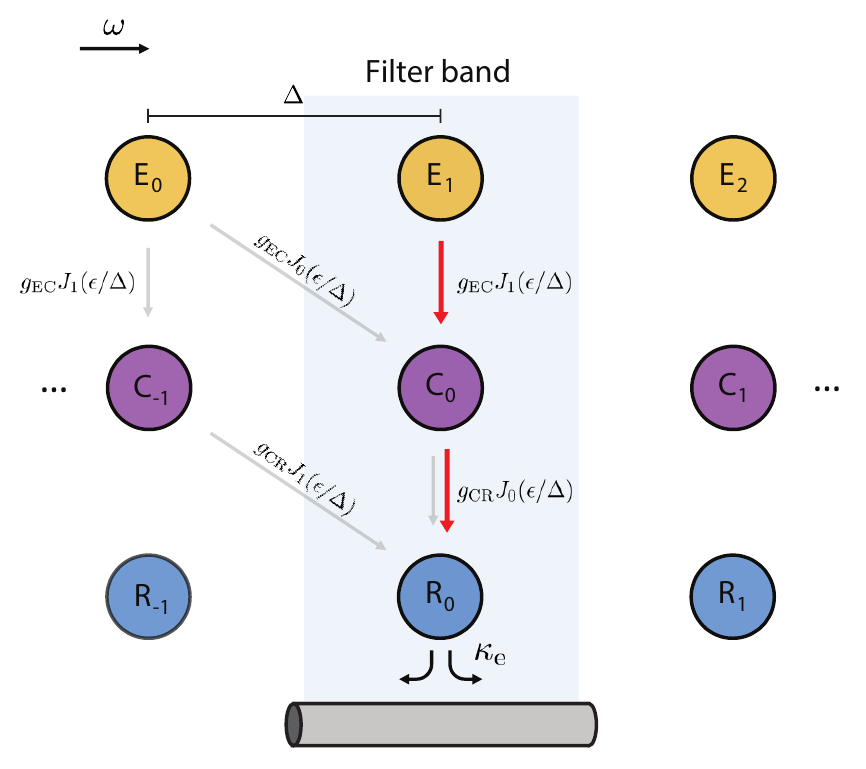}
\caption{\textbf{Decay pathways of emitter sidebands} The filter cavity at its natural frequency R$_0$ is the source of waveguide decay. The first emitter sideband's (E$_1$) dominant decay pathway occurs through interaction with the coupler baseband and cavity baseband (red arrows, E$_0$ $\rightarrow$ C$_{1}$ $\rightarrow$ R$_0$). Decay pathways of the parasitic emitter baseband (E$_0$) are suppressed by the filter cavity (gray arrows). Note that the emitter sideband spectrum is offset from the cavity and coupler sideband spectra by $\Delta$, corresponding to the experimental settings.}
\label{fig:filter}
\end{figure}

The derived input-output relations provide a method for estimating the external coupling of the emitter and its sidebands. In principle, there are infinitely many decay pathways for a given emitter sideband; each pathway is mediated by a coupler sideband of distinct order. This is evident in the mapping of interaction strengths given in \cref{fig:CMT_theory}b. In this section, we will make estimates for the waveguide coupling rate of an emitter sideband E$_{n}$ arising from interaction with a coupler sideband C$_{m}$; the full decay pathway is E$_{n}$ $\rightarrow$ C$_{m}$ $\rightarrow$ R$_0$. We consider three cases, the dispersive regime, hybridized coupler-filter cavity regime, and fully hybridized regime. We will then apply these estimates to illustrate waveguide decay pathways in our experiment.

For the radiation pathway E$_n$ $\rightarrow$ C$_m$ $\rightarrow$ R$_0$, relevant frequencies are $\omega_{\mathrm{E}n} = \omega_\mathrm{E}+n\Delta$, $\omega_{\mathrm{C}m} = \omega_\mathrm{C}+m\Delta$, and $\omega_\mathrm{R}$. We will refer to these relevant sidebands as the emitter, coupler, and filter cavity.

\subsubsection{Dispersive regime}
When the coupler is far detuned from the emitter and filter cavity ($\delta_{\mathrm{C}m,\mathrm{E}n} \gg g_\mathrm{EC}J_{m-n}({\epsilon}/{\Delta}),  \delta_{\mathrm{C}m,\mathrm{R}} \gg g_\mathrm{CR}J_m({\epsilon}/{\Delta})$), the effective interaction strength between the emitter and filter cavity is given by
\begin{equation}
    g_\mathrm{eff} = \frac{g_\mathrm{EC} g_\mathrm{CR}}{2} J_{m-n}(\frac{\epsilon}{\Delta})  J_m(\frac{\epsilon}{\Delta}) (\frac{1}{\delta_{\mathrm{C}m,\mathrm{E}n}}+\frac{1}{\delta_{\mathrm{C}m,\mathrm{R}}})
\end{equation}
under the rotating wave approximation \cite{Yan2018Nov}. Provided that $\delta_{\mathrm{E}n,\mathrm{R}} \gg g_\mathrm{eff}$, where $\delta_{\mathrm{E}n,\mathrm{R}} = \omega_{\mathrm{E}n} - \omega_\mathrm{R}$, emitter external coupling to the waveguide is then given by

\begin{equation}
    \kappa_\mathrm{em} = (\frac{g_\mathrm{eff}}{\delta_{\mathrm{E}n,\mathrm{R}}})^2\kappa_\mathrm{e} 
\end{equation}

\subsubsection{Hybridized coupler-filter cavity regime} 
The coupler and filter cavity may be strongly hybridized with the emitter far detuned from either of the hybrid modes. In this situation, $g_\mathrm{eff}$ between the emitter and the hybridized mode (i.e., the majority coupler hybridized mode) can be expressed as

\begin{equation}
    g_\mathrm{eff} = g_\mathrm{EC} J_{m-n}(\frac{\epsilon}{\Delta}) (\frac{\zeta^2}{\xi^2+\zeta^2})
    \label{eq:geff_hybrid}
\end{equation}

where the raising operator for the hybridized mode is
\begin{equation}
    \hat{a}^\dagger_\mathrm{h} = \frac{1}{\sqrt{\zeta^2+\xi^2}}(\zeta \hat{a}^\dagger_\mathrm{C}+\xi \hat{a}^\dagger_\mathrm{R})
\end{equation}

and has frequency $\omega_\mathrm{h}$. With $\delta_{\mathrm{E}n,h} = \omega_{\mathrm{E}n} - \omega_\mathrm{h}$, Emitter external coupling to the waveguide is then given by

\begin{equation}
    \kappa_\mathrm{em} = (\frac{g_\mathrm{eff}}{\delta_{\mathrm{E}n,\mathrm{h}}})^2(\frac{\xi^2}{\xi^2+\zeta^2})\kappa_\mathrm{e} 
\end{equation}

\subsubsection{Fully hybridized regime} In the case of strong hybridization of the coupler and filter cavity, we may also allow the emitter sideband to hybridize with a coupler-filter cavity hybrid mode. The emitter's raising operator then becomes

\begin{equation}
    \hat{a}^\dagger_\mathrm{d} = \frac{1}{\sqrt{1+\alpha^2}}(\hat{a}^\dagger_\mathrm{E}+\alpha\hat{a}^\dagger_\mathrm{h})
\end{equation}

where $\alpha \ll 1$. External coupling is then given by the following.
\begin{equation}
    \kappa_\mathrm{em} = (\frac{\alpha^2}{1+\alpha^2})(\frac{\xi^2}{\xi^2+\zeta^2})\kappa_\mathrm{e} 
\end{equation}

Generally, for a fixed drive amplitude ($\epsilon$) and drive frequency ($\Delta$), permitting stronger emitter hybridization with the coupler and resonator results in larger emitter-waveguide coupling. Hence, full hybridization  yields larger emitter-waveguide coupling than only hybridizing the coupler and filter cavity. Operating in the dispersive regime results in the lowest emitter-waveguide coupling.

\subsubsection{Chiral emitter sideband decay}

In our application, we are primarily concerned with improving the external coupling of the emitter qubit's first sideband E$_1$. Our experimental settings correspond to modest values of ${\epsilon}/{\Delta}<1$, so that only interactions between adjacent sidebands are significant ($J_{n>2}(\epsilon/\Delta) \approx 0$). Hence, the potential decay pathways for E$_1$ reduce to either E$_1$ $\rightarrow$ C$_0$ $\rightarrow$ R$_0$ or E$_1$ $\rightarrow$ C$_1$ $\rightarrow$ R$_0$. 

Our device is designed to maximize decay through the first channel (E$_1$ $\rightarrow$ C$_0$ $\rightarrow$ R$_0$), marked by red arrows in \cref{fig:filter}. Because $\Delta > g_\mathrm{CE}$, $g_\mathrm{CR}$ in our device, this may be done by positioning the emitter sideband and coupler baseband frequencies near the cavity baseband frequency ($\delta_\mathrm{C,E1} \approx g_\mathrm{EC}, \delta_\mathrm{C,R} \approx g_\mathrm{CR} $). These frequency spacings allow for the hybridization of the coupler and cavity basebands. As a result, this dominant decay pathway operates in the (b) hybridized coupler-filter cavity regime.

Under these conditions, decay through the second channel (E$_1$ $\rightarrow$ C$_1$ $\rightarrow$ R$_0$) is suppressed due to detuning between C$_1$ and R$_0$ (of approximately $\Delta$). The second channel operates in the (a) dispersive regime.

\subsubsection{Parasitic emitter sideband decay}

We now consider the decay of the emitter baseband (E$_0$) to illustrate the spectral filtering provided by the cavities. Again, because experiments operate within ${\epsilon}/{\Delta}<1$, we consider only interactions between adjacent sidebands. The relevant decay pathways are then  E$_0$ $\rightarrow$ C$_{1}$ $\rightarrow$ R$_0$, E$_0$ $\rightarrow$ C$_0$ $\rightarrow$ R$_0$, and E$_0$ $\rightarrow$ C$_{-1}$ $\rightarrow$ R$_0$. As shown in \cref{fig:filter}, the first pathway E$_0$ $\rightarrow$ C$_{1}$ $\rightarrow$ R$_0$, operates strongly in the (a) dispersive regime. The second channel, E$_0$ $\rightarrow$ C$_0$ $\rightarrow$ R$_0$ (marked in gray arrows in \cref{fig:filter}]), operates in the (b) hybridized coupler-filter cavity regime. In this case, the detuning between the emitter baseband and the hybridized coupler-filter cavity suppresses the external coupling. The last channel, E$_0$ $\rightarrow$ C$_{-1}$ $\rightarrow$ R$_0$ (marked in gray arrows in \cref{fig:filter}), is not described by the three operating regimes presented. In this pathway the emitter decay is suppressed by the large detuning between the coupler sideband C$_{-1}$ and filter cavity.

\subsection{Experimental operation of tunable coupler}

\begin{figure}
\centering
\includegraphics[width=0.9\linewidth]{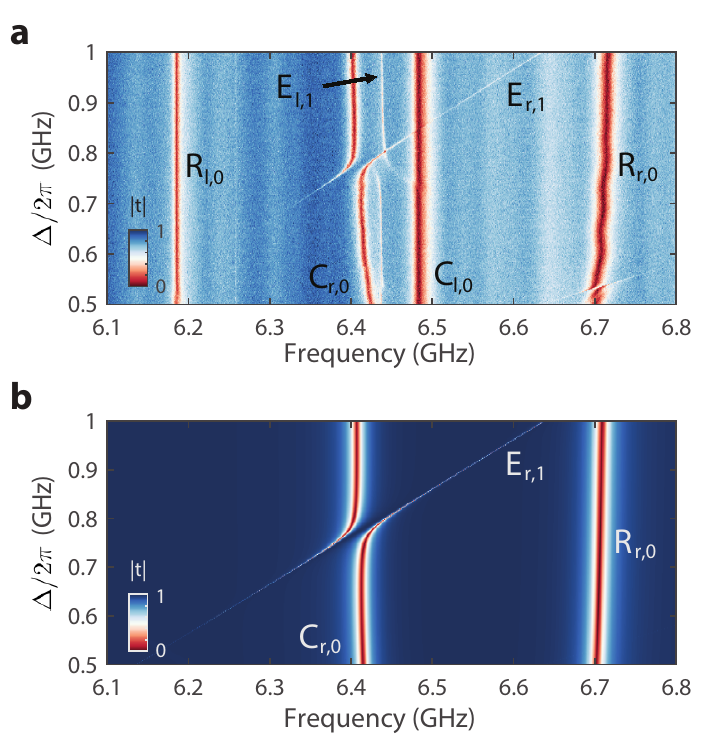}
\caption{\textbf{Comparison of experimental transmission and input-output calculation} (a) Measured device transmission as RF drive frequency of the right coupler is swept from 500 MHz to 1 GHz. The left coupler drive frequency is set to 805 MHz during the entire measurement. The emitter's first blue sideband frequency (generated by the right coupler) increases with $\Delta/2\pi$. The same sideband, generated by the left coupler, remains at 6.441 GHz. Spectrally overlapping these sidebands yields the interference required for chirality. (b) Calculated device transmission for the right coupler. The emitter's first blue sideband, coupler baseband, and filter cavity baseband are visible.}
\label{fig:CMT_calibration}
\end{figure}

To operate the tunable coupler, we first DC bias the coupler qubit to an appropriate working point, as shown in \cref{fig:1}c (bottom panel). We then apply a microwave tone to the coupler qubit flux line (Z$_\mathrm{l}, \mathrm{Z}_\mathrm{r}$ shown in \cref{fig:1}b). Because the coupler qubit has a non-linear dependence on the flux threading its SQUID loop (shown in \cref{fig:tuning_curves}a,b), the amplitude $\epsilon$ and coupler mean frequency $\omega_\mathrm{C}$ are determined by both the DC bias point ($\Phi_\mathrm{DC}$) and the applied RF power ($P_\mathrm{in} \propto \epsilon^2$). Taylor expanding the coupler qubit frequency results in the following relations. 

\begin{equation}
    \omega_\mathrm{C}(\Phi(t)) = \omega_\mathrm{C}(\Phi_\mathrm{DC}+\epsilon_{\Phi}\text{sin}(\Delta t)) \approx \bar{\omega}_\mathrm{C} + \epsilon \text{sin}(\Delta t)
\end{equation}

where $\bar{\omega}_\mathrm{C} = \omega_\mathrm{C} + \frac{\epsilon_{\Phi}^2}{4}\frac{d\omega_\mathrm{C}}{d\Phi}\Bigr|_{\substack{\Phi_\mathrm{DC}}}$ and $\epsilon = \epsilon_\Phi\frac{d\omega_\mathrm{C}}{d\Phi}\Bigr|_{\substack{\Phi_\mathrm{DC}}}$. Note the change in the notation of $\bar{\omega}_\mathrm{C}$ to represent the coupler mean frequency (given by ${\omega}_\mathrm{C}$ in previous sections). 

In \cref{fig:CMT_calibration}, we use the input-output model to reproduce the qualitative behavior of the tunable coupler. In \cref{fig:CMT_calibration}a, we record waveguide transmission while sweeping the RF drive frequency of the right coupler qubit. All qubits are set to the same DC flux bias as the chiral configuration presented in \cref{fig:1}c. Additionally, the RF drive frequency of the left coupler qubit is set to $\Delta_\mathrm{l}/2\pi$ = 805 MHz. As a result, the $\Delta_\mathrm{r}/2\pi$ = 805 MHz transmission trace in \cref{fig:CMT_calibration}(a) corresponds to \cref{fig:1}c.

\cref{fig:CMT_calibration}(b) gives the waveguide transmission calculated using the input-output model, including only the right decay pathway. As a result, the left coupler, filter cavity, and left emitter sideband are not present. 

The calculated transmission uses experimentally extracted parameters for the emitter, coupler, and filter cavity (given in \cref{tab:table1}, \cref{tab:table2}, \cref{tab:table3}). The drive amplitude $\epsilon$ is calibrated by recording the shift in the coupler qubit mean frequency while sweeping the RF drive power. For $\Delta/2\pi = $ 805 MHz, drive amplitude is $\epsilon/2\pi = $ 364MHz, yielding $\epsilon/\Delta = $ 0.452. At this drive amplitude, $J_0({\epsilon}/{\Delta}) =$ 0.95, $J_1({\epsilon}/{\Delta}) =$ 0.22, and $J_2({\epsilon}/{\Delta}) =$ 0.03. The Bessel functions of higher order ($>$ 2) may therefore be safely ignored, and the input-output matrix equation is truncated after the second-order negative and positive sidebands ($\bf{H_{0}}$, $\bf{H_{\pm 1}}$, $\bf{H_{\pm 2}}$ are included).

\end{document}